\documentclass[aps,prd,amsmath,amssymb,showpacs,10pt,superscriptaddress]{revtex4-2}
\usepackage{graphicx}
\usepackage{color}
\usepackage{slashed}
\usepackage{subfigure}
\usepackage{enumerate}
\usepackage{physics}
\usepackage[font=footnotesize]{caption}
\usepackage[breaklinks,colorlinks,urlcolor=blue,linkcolor=red,anchorcolor=magenta,citecolor=blue]{hyperref}
\usepackage{cleveref} 
\crefname{table}{Table}{Tables}
\crefname{equation}{Eq.}{Eqs.}
\crefname{figure}{Fig.}{Figs.}
\crefname{section}{Sec.}{Secs.}
\allowdisplaybreaks
\begin{document}
\title{Predictions for the isospin-violating decays of $B_{c}(1P)^{+}\to B_{c}^{(*)+}\pi^{0}$}
\author{Jun Wang
}\email{junwang@ihep.ac.cn}
\affiliation{Institute of High Energy Physics, Chinese Academy of Sciences, Beijing 100049, P.R. China}
\affiliation{University of Chinese Academy of Sciences, Beijing 100049, P.R. China}
\author{Qiang Zhao
}\email{ zhaoq@ihep.ac.cn} 
\affiliation{Institute of High Energy Physics, Chinese Academy of Sciences, Beijing 100049, P.R. China}
\affiliation{University of Chinese Academy of Sciences, Beijing 100049, P.R. China}
\affiliation{Center for High Energy Physics, Henan Academy of Sciences, Zhengzhou 450046, P.R. China}

\begin{abstract}
In this work we study the isospin-violating decays of  $B_{c}(1P)^{+}\to B_{c}^{(*)+}\pi^{0}$, which may provide additional information for the determination of the properties of the first orbital excitation states of $B_{c}(1P)^{+}$. By assuming a dual relation between the U(1) anomaly soft-gluon coupling for $B_{c}(1P)^{+}\to B_{c}^{(*)+}\pi^{0}$ and the intermediate meson loop transitions, we can quantify the isospin-violating decay effects for these four $P$-wave states. We find that the partial decay width of $B_{c0}^{*+}\to B_{c}^{+}\pi^{0}$ is about three orders of magnitude larger than that for $B_{c2}^{*+}\to B_{c}^{+}\pi^{0}$. It implies that $B_{c0}^{*+}$ can be established in the $B_{c}^{+}\pi^{0}$ decay channel as a single state. Meanwhile, the two axial-vector states $B_{c1}^{+}/B_{c1}'^{+}$ can be possibly identified in $B_{c1}^{+}/B_{c1}'^{+}\to B_{c}^{*+}\pi^{0}$ with comparable strengths. Although these isospin-violating decays turn out to be small, the theoretical predictions should be useful for guiding future experimental efforts. 
\end{abstract}

\maketitle

\section{Introduction}\label{sec:intro}
As the only meson containing two different heavy quarks, the $B_c$ mesons family provides a unique perspective for studying the properties of the non-perturbative regime of quantum chromodynamics (QCD). The mass spectrum of $B_c$ mesons has been predicted by many theoretical approaches~\cite{Gupta:1995ps,Li:2023wgq,Li:2022bre,Li:2019tbn,Eichten:2019gig,Ebert:2002pp,Fulcher:1998ka,Gershtein:1994dxw,Wang:2022cxy,Davies:1996gi,Hao:2024nqb,Godfrey:2016nwn,Godfrey:2004ya}. Experimentally, however, only the ground state $B_c^+$~\cite{CDF:1998ihx} and the first radial excitation $B_c(2S)^+$~\cite{LHCb:2017rqe,ATLAS:2014lga,CMS:2019uhm,LHCb:2019bem} have been observed so far. In addition, the mass and decay modes of the $B_c^+$ have been measured with high precision~\cite{LHCb:2014mvo,LHCb:2019tea,LHCb:2014glo,LHCb:2014ilr,LHCb:2022htj,LHCb:2020ayi,LHCb:2024nlg,LHCb:2016djy}. However, for excited $B_c$ states beyond the ground state, experimental information on their decay modes remains scarce.

Recently, the first observation of the $B_c(1P)^+$, the first orbital excitation of the $B_c$ meson, was reported by the LHCb Collaboration in the invariant mass spectrum of the $B_c^+\gamma$ final state~\cite{LHCb:2025ubr,LHCb:2025uce}. The $B_c(1P)^+$ multiplet consists of four physical states: $B_{c0}^{*+}(1^{3}P_{0})$, $B_{c1}^{+}(1P_{1})$, $B_{c1}'(1P'_{1})$, and $B_{c2}^{*+}(1^{3}P_{2})$, where $1P_{1}$ and $1P'_{1}$ are mixtures of the $1^{1}P_{1}$ and $1^{3}P_{1}$ states~\cite{Godfrey:2004ya}. All these four states can decay into $B_c^{*+}\gamma$ with $B_c^{*+}\to B_c^+\gamma$ to be detected in the final state. Meanwhile, the $1P_{1}$ and $1P'_{1}$ states can reach the $B_c^+\gamma$ final state directly through an $S$-wave transition. However, due to the limited statistics and energy resolution, only one photon associated with the observed two peaks in the $B_c^+\gamma$ spectrum are actually effective. Namely, LHCb only measured a $B_c^+$ plus a photon in the final state. Thus, it is likely that the data actually contain contributions from all these four states.

In addition to the radiative decays, by analogy with the isospin-violating decays of $D_{s}^{*+}$~\cite{Wang:2025fzj,Yang:2019cat}, the $B_{c}(1P)^{+}$ states can also decay to the final states $B_{c}^{(*)+}\pi^{0}$ via isospin-violating processes ($B_{c0}^{*}/B_{c2}^{*}\to B_{c}^{+}\pi^{0}$, $B_{c1}^{+}/B_{c1}'^{+}\to B_{c}^{*+}\pi^{0}$). Although the $B_{c}^{*+}$ has not yet been observed experimentally, it is rather reliable that the $B_{c}^{*+}$ mass is below the $B_{c}^{+}\pi^{0}$ threshold through theoretical studies~\cite{Godfrey:2004ya}. Therefore, the decays of $B_{c0}^{*}/B_{c2}^{*}\to B_{c}^{+}\pi^{0}$ and $B_{c1}^{+}/B_{c1}'^{+}\to B_{c}^{*+}\pi^{0}$ can provide more information about these four $P$-wave states. 

In this work, we employ the effective Lagrangian approach to calculate the decay widths of the isospin-violating processes $B_{c}(1P)^{+}\to B_{c}^{(*)+}\pi^{0}$. 
At the quark level the isospin-violating decay $B_{c}(1P)^{+}\to B_{c}^{(*)+}\pi^{0}$ can be described by the soft-gluon coupling to the light $u\bar{u}$ and $d\bar{d}$ via the U(1) anomaly term. Meanwhile, since $\pi^0$ is an isovector with the flavor wavefunction $(u\bar{u}-d\bar{d})/\sqrt{2}$, the two terms will cancel each other. Taking into account the property of the U(1) anomaly coupling which is proportional to the quark mass and only partially conserves the axial current, the isospin-violating coupling is proportional to a term of $m_d-m_u$, which will be convoluted by the transition form factor for  $B_{c}(1P)^{+}\to B_{c}^{(*)+}$. 
According to the concept of quark-hadron duality~\cite{Shifman:2000jv}, this process can be interpreted as the $B_{c}(1P)^{+}$ meson producing intermediate $\mathcal{B}^{(*)}$ and $\mathcal{D}^{(*)}$ states, which then rescatterings via the exchange of a $\mathcal{D}^{(*)}$ meson to produce the $B_{c}^{(*)+}\pi^{0}$ final state. The two loop amplitudes involving the intermediate mesons with $u$ and $d$ can have constructive phase to produce $\eta$ meson or destructive phase to produce $\pi^0$ due to isospin conservation. This mechanism can be quantified by the effective Lagrangian approach, where the vertex couplings can be determined by the $^{3}P_{0}$ model~\cite{LeYaouanc:1972vsx,LeYaouanc:1988fx,Blundell:1996as}. This method has been broadly applied to the studies of various Okubo-Zweig-Iizuka (OZI) rule~\cite{Okubo:1963fa} evading processes~\cite{Zhang:2009kr,Zhang:2009gy,Guo:2010ak}, isospin symmetry breaking processes~\cite{Li:2007au,Li:2008xm,Wang:2011yh,Guo:2010ak,Guo:2010zk,Wang:2025fzj} and the helicity selection rule violation processes~\cite{Liu:2009vv,Liu:2010um,Zhang:2009kr} in the literature. 

As follows, We first introduce our formalism in \cref{sec:formalism}. The numerical results and discussions are presented in \cref{sec:results}. A brief summary and conclusion are given in \cref{sec:summary}. 

\section{Formalism}\label{sec:formalism}

In this section we first present the effective Lagrangians, and derive the loop amplitudes for $B_{c}(1P)^{+}\to B_{c}^{(*)+}\pi^{0}$. We then determine the coupling constants appearing in the effective Lagrangians within the framework of the $^{3}P_{0}$ model~\cite{LeYaouanc:1972vsx}.

\subsection{Effective Lagrangians}
The effective Lagrangians for the $B_{cJ} $ couplings to $\mathcal{B}^{(*)}\mathcal{D}^{(*)}$ are as follows~\cite{Zou:2002ar,Dulat:2005in}:
\begin{align}
\mathcal{L}_{B_{c}}=&ig_{B_{c}\mathcal{B}^{*}\mathcal{D}}B_{c}\mathcal{B}^{*i}_{\mu}\partial^\mu \mathcal{D}_{i}+ig_{B_{c}\mathcal{B}\mathcal{D}^{*}}B_{c}\partial^\mu\mathcal{B}_{i} \mathcal{D}^{*i}_{\mu}+g_{B_{c}\mathcal{B}^{*}\mathcal{D}^{*}}\varepsilon_{\mu\nu \rho\sigma}B_{c}\partial^{\mu}\mathcal{B}^{*i\nu}\partial^{\rho}\mathcal{D}^{*\sigma}_{i}+h.c.\,,\\
  \mathcal{L}_{B_{c0}^{*}}=&g_{B_{c0}^{*}\mathcal{B} \mathcal{D}}B_{c0}^{*}\mathcal{B}^{i}\mathcal{D}_{i}+g_{B_{c0}^{*}\mathcal{B}^{*}\mathcal{D}^{*}}B_{c0}^{*}\mathcal{B}^{*i}_{\mu}\mathcal{D}_{i}^{*\mu}+h.c.\,,\\ 
  \mathcal{L}_{B_{c2}^{*}}=& g_{B_{c2}^{*}\mathcal{B} \mathcal{D}}\partial_{\mu}\mathcal{B}^{i} \partial_{\nu}\mathcal{D}_{i}B_{c2}^{*\mu \nu}+g_{B_{c2}^{*}\mathcal{B}^{*} \mathcal{D}^{*}}B^{*\mu\nu}_{c2}\mathcal{B}^{*i}_{\mu}\mathcal{D}^{*}_{i\nu}+h.c.\,,\\
   \mathcal{L}_{B_{c}^{*}}=& ig_{B_{c}^{*}\mathcal{B}\mathcal{D}}B_{c}^{*\mu}\partial_\mu \mathcal{B}_{i}\mathcal{D}^{i}+ g_{B_{c}^{*}\mathcal{B}^{*}\mathcal{D}}\varepsilon_{\mu\nu\rho\sigma}\partial^{\mu}B_{c}^{*\nu}\partial^{\rho}\mathcal{B}^{*\sigma}_{i}\mathcal{D}^{i}+ g_{B_{c}^{*}\mathcal{B}\mathcal{D}^{*}}\varepsilon_{\mu\nu\rho\sigma}\partial^{\mu}B_{c}^{*\nu}\partial^{\rho}\mathcal{D}^{*\sigma}_{i}\mathcal{B}^{i}\nonumber\\
   &+ig_{B_{c}^{*}\mathcal{B}^{*}\mathcal{D}^{*}}(\partial^\mu B_{c}^{*\nu}-\partial^{\nu}B_{c}^{*\mu})\mathcal{B}^{*}_{i\mu}\mathcal{D}^{*i}_{\nu}+h.c.\,,\\
   \mathcal{L}_{B_{c1}}=& g_{B_{c1}\mathcal{B}\mathcal{D}^{*}}B_{c1}^{\mu}\mathcal{D}^{*i}_{\mu}\mathcal{B}_{i}+g_{B_{c1}\mathcal{B}^{*}\mathcal{D}}B_{c1}^{\mu}\mathcal{B}^{*i}_{\mu}\mathcal{D}_{i}+ig_{B_{c1}\mathcal{B}^{*}\mathcal{D}^{*}}\varepsilon_{\mu\nu \rho\sigma}\partial^{\mu}B_{c1}^{\nu}\mathcal{B}^{*\rho}_{i}\mathcal{D}^{*i\sigma}+h.c.\label{eq:eflbc} \ , \\
      \mathcal{L}_{B_{c1}^{\prime}}=& g_{B_{c1}^{\prime}\mathcal{B}\mathcal{D}^{*}}B_{c1}^{\prime\mu}\mathcal{D}^{*i}_{\mu}\mathcal{B}_{i}+g_{B_{c1}^{\prime}\mathcal{B}^{*}\mathcal{D}}B_{c1}^{\prime\mu}\mathcal{B}^{*i}_{\mu}\mathcal{D}_{i}+ig_{B_{c1}^{\prime}\mathcal{B}^{*}\mathcal{D}^{*}}\varepsilon_{\mu\nu \rho\sigma}\partial^{\mu}B_{c1}^{\prime\nu}\mathcal{B}^{*\rho}_{i}\mathcal{D}^{*i\sigma}+h.c. \ ,
   \end{align}
where $\mathcal{B}=(B^{+},B^{0},B_{s}^{0})$ and $\mathcal{D}=(D^{0},D^{+},D^{+}_{s})$. The coupling constants for the $B_{c(0,1,2)}^{(*)}\mathcal{B}^{(*)}\mathcal{D}$ vertices will be calculated using the $^{3}P_{0}$ model~\cite{Ackleh:1996yt,LeYaouanc:1972vsx}.

The effective Lagrangian for the $\mathcal{D}^{(*)}\mathcal{D}^{(*)}\mathcal{P}$ coupling is given by~\cite{Casalbuoni:1996pg,Cheng:2004ru}
\begin{equation}
    {\cal L}_{\mathcal{D}^{(*)}\mathcal{D}^{(*)}\mathcal{P}} = - ig_{\mathcal{D}^*\mathcal{D}\mathcal{P}}(\mathcal{D}^i\partial^\mu \mathcal{P}_{ij}
    \mathcal{D}_\mu^{*j\dagger}-\mathcal{D}_\mu^{*i}\partial^\mu \mathcal{P}_{ij}\mathcal{D}^{j\dagger})
    +\frac{1}{2}g_{\mathcal{D}^*\mathcal{D}^*\mathcal{P}}
\varepsilon_{\mu\nu\alpha\beta}\,\mathcal{D}_i^{*\mu}\partial^\nu \mathcal{P}^{ij}
    {\overleftrightarrow \partial}{}^{\!\alpha} \mathcal{D}^{*\beta\dagger}_j \,,
 \end{equation} 
 where
 \begin{equation}
   \mathcal{P}\equiv
\begin{pmatrix}
    \frac{\sin \alpha_{P}\eta^{\prime}+\cos \alpha_P \eta +\pi^{0}}{\sqrt{2}}& \pi^+ & K^+ \\
    \pi^- &\frac{\sin \alpha_{P}\eta^{\prime}+\cos \alpha_P \eta -\pi^{0}}{\sqrt{2}}  & K^0 \\[1ex]
    K^-& {\bar K}^0 &\cos  \alpha_{P}\eta^{\prime}-\sin  \alpha_P \eta \\
\end{pmatrix}\,,
  \end{equation} 
with $\alpha_P=40.6^\circ$ adopted for the $\eta-\eta^{\prime}$ mixing angle in the SU(3) flavor basis, and the value is an average from the Particle Data Group~\cite{ParticleDataGroup:2024cfk}.

The coupling constants for the effective interaction vertices $\mathcal{D}^{(*)}\mathcal{D}^{(*)}\pi^{0}$ can be obtained from heavy quark effective theory~\cite{Casalbuoni:1996pg,Cheng:2004ru,Wang:2012mf,Wang:2011yh}:
\begin{equation}  g_{\mathcal{D}^{*}\mathcal{D}^{*}\pi}=\frac{g_{\mathcal{D}^{*}\mathcal{D}\pi}}{\sqrt{m_{\mathcal{D}}m_{\mathcal{D}^{*}}}}=\frac{2}{f_{\pi}}g,\quad g_{D^{(*)}_{s}D^{(*)}_{s}\mathcal{P}}=\sqrt{\frac{m_{D_{s}^{(*)}}m_{D_{s}^{(*)}}}{m_{D^{(*)}}m_{D^{(*)}}}}g_{\mathcal{D}^{(*)}\mathcal{D}^{(*)}\pi} \label{eq:gddp}
 \end{equation} 
where $g=0.59$, and $f_{\pi} = 132\,\mathrm{MeV}$ is the $\pi$ decay constant and $m_{\rho}$ is the mass of the $\rho$ meson.

Due to the $\eta-\pi^{0}$ mixing effect,  we need to consider the contribution from the isospin-conserving transition $\mathcal{D}^{(*)}\to \mathcal{D}^{(*)}\eta$ followed by $\eta \to \pi^{0}$ to violate the isospin symmetry.
   For the $\eta \to \pi^{0}$ process, we use the $\eta-\pi^{0}$ mixing angle $\theta_{\eta \pi^{0}}$ which is given by the leading order chiral expansion~\cite{Gasser:1984gg}
     \begin{equation}
       \tan(2\theta_{\eta \pi^0})=\frac{\sqrt{3}}{2}\frac{m_{d}-m_{u}}{m_{s}-\hat{m}}.
      \end{equation} 
where $\hat{m}=(m_{u}+m_{d})/2$. The values of $m_{u}$, $m_{d}$, and $m_{s}$ are taken from the PDG~\cite{ParticleDataGroup:2024cfk}. Since $\theta_{\eta \pi^0}$ is very small, we take
\begin{equation}
       \theta_{\eta \pi^0}\simeq \frac{\sqrt{3}}{4}\frac{m_{d}-m_{u}}{m_{s}-\hat{m}},
\end{equation}
as broadly adopted in the literature.
\subsection{Loop transition amplitudes of $B_{c}(1P)^{+}\to B_{c}^{(*)+}\pi^{0}$}
Based on the effective Lagrangians given above, we can derive the corresponding loop amplitudes. 
\begin{figure}[htbp]
\centering 
\subfigtopskip=2pt \subfigbottomskip=2pt \subfigcapskip=-10pt 
\subfigure[]{\includegraphics[width=0.32\textwidth]{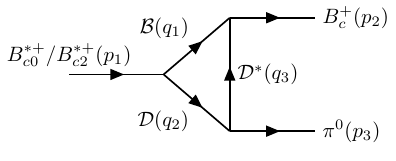}}
\subfigure[]{\includegraphics[width=0.32\textwidth]{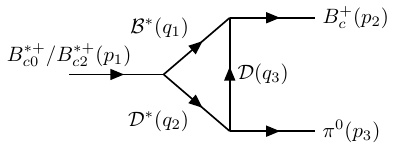}}
\subfigure[]{\includegraphics[width=0.32\textwidth]{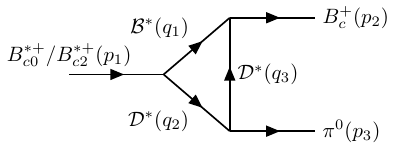}}
\subfigure[]{\includegraphics[width=0.32\textwidth]{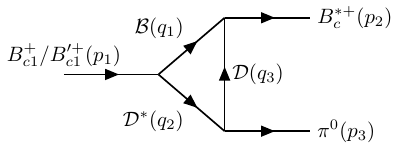}}
\subfigure[]{\includegraphics[width=0.32\textwidth]{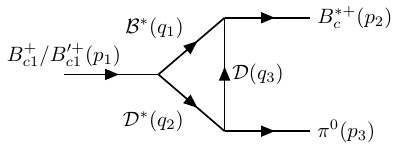}}
\subfigure[]{\includegraphics[width=0.32\textwidth]{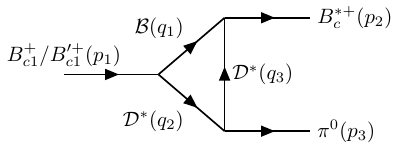}}
\subfigure[]{\includegraphics[width=0.32\textwidth]{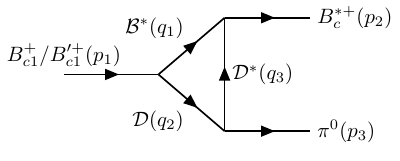}}
\subfigure[]{\includegraphics[width=0.32\textwidth]{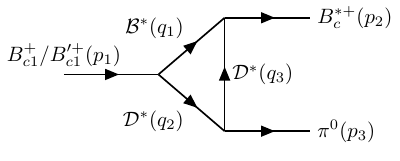}}
\caption{Schematic diagrams of the decay $B_{c}(1P)^{+}\to B_{c}^{(*)+}\pi^{0}$ via intermediate meson loops without $\eta-\pi^{0}$ mixing, where $\mathcal{B}=(B^{+},B^{0})$ and $\mathcal{D}=(D^{0},D^{+})$. (a)--(h) illustrate different intermediate meson loops.}\label{fig:pi}
\end{figure}

For the loop-level amplitudes without the $\eta-\pi^{0}$ mixing corresponding to Fig.\,\ref{fig:pi}, we denote the amplitude as $i\mathcal{M}_k(P_1,P_2,P_3)$, where $k=0,2,1$ and $1^{\prime}$ corresponds to $B_{c0}^{*+},B_{c2}^{*+},B_{c1}^{+}$ and $B_{c1}^{\prime+}$, respectively. Here, $P_i$ represents the intermediate meson with momentum $q_i$ and $B_{\mu\nu}$ denotes the polarization tensor of the initial $B_{c2}^{*+}$.  $\varepsilon^{i}_{\mu}$ denotes the polarization vector of the initial $B_{c1}^{(\prime)+}$ and $\varepsilon^{f}_{\mu}$ denotes the polarization vector of the final $B_{c}^{*+}$.  The amplitudes corresponding to Fig.\,\ref{fig:pi} can thus be expressed as
\begin{align}
  i \mathcal{M}_{0}(\mathcal{B},\mathcal{D},\mathcal{D}^{*})=& \int \frac{d^{4}q_{3}}{(2\pi)^{4}} \frac{g_{B^{*}_{c0}\mathcal{B}\mathcal{D}}g_{B_{c}\mathcal{B}\mathcal{D}^{*}}g_{\mathcal{D}^{*}\mathcal{D}\mathcal{P}}q_{1}^{\mu}\qty(g_{\mu\nu}-\frac{q_{3\mu}q_{3\nu}}{m^{2}_{3}})p_{3}^{\nu}}{(q_1^{2}-m_1^{2})(q_2^{2}-m_2^{2})(q_3^{2}-m_3^{2})}\mathcal{F}(q_{i}^{2})\,,\\
  i \mathcal{M}_{0}(\mathcal{B}^{*},\mathcal{D}^{*},\mathcal{D})=&\int \frac{d^{4}q_{3}}{(2\pi)^{4}} \frac{g_{B_{c0}^{*}\mathcal{B}^{*}\mathcal{D}^{*}}g_{B_{c}\mathcal{B}^{*}\mathcal{D}}g_{\mathcal{D}^{*}\mathcal{D}\mathcal{P}}\qty(g^{\mu\nu}-\frac{q_{1}^{\mu}q_{1}^{\nu}}{m_1^{2}})\qty(g_{\mu \alpha}-\frac{q_{2\mu}q_{2\alpha}}{m_2^{2}})q_{3\nu}p_{3}^{\alpha}}{(q_1^{2}-m_1^{2})(q_2^{2}-m_2^{2})(q_3^{2}-m_3^{2})}\mathcal{F}(q_{i}^{2})  \,,\\
   i \mathcal{M}_{0}(\mathcal{B}^{*},\mathcal{D}^{*},\mathcal{D}^{*}) =&\int \frac{d^{4}q_{3}}{(2\pi)^{4}} \frac{ g_{B_{c0}^{*}\mathcal{B}^{*} \mathcal{D}^{*}} g_{B_{c}\mathcal{B}^{*}\mathcal{D}^{*}}g_{\mathcal{D}^{*}\mathcal{D}^{*}\mathcal{P}}\qty(g^{\mu\nu}-\frac{q_{1}^{\mu}q_{1}^{\nu}}{m_1^{2}})g_{\mu\alpha}\qty(g^{ \alpha \beta}-\frac{q_{2}^{\alpha}q_{2}^{\beta}}{m_2^{2}})}{(q_1^{2}-m_1^{2})(q_2^{2}-m_2^{2})(q_3^{2}-m_3^{2})}\\&\times \qty(g^{\rho\sigma}-\frac{q_{3}^{\rho}q_{3}^{\sigma}}{m^{2}_{3}})\varepsilon_{\nu \delta \rho \kappa}q_{1}^{\delta}q_{3}^{\kappa} \varepsilon_{\beta \lambda \sigma\tau}q_{3}^{\lambda}q_{2}^{\tau}\mathcal{F}(q_{i}^{2}) \,, \\
  i \mathcal{M}_{2}(\mathcal{B},\mathcal{D},\mathcal{D}^{*})=&\int \frac{d^{4}q_{3}}{(2\pi)^{4}} \frac{g_{B^{*}_{c2}\mathcal{B}\mathcal{D}}g_{B_{c}\mathcal{B}\mathcal{D}^{*}}g_{\mathcal{D}^{*}\mathcal{D}\mathcal{P}}B_{\alpha \beta}q_1^{\alpha}q_{2}^{\beta}\qty(g_{\mu\nu}-\frac{q_{3\mu}q_{3\nu}}{m^{2}_{3}}) q_{1}^{\mu}p_{3}^{\nu}}{(q_1^{2}-m_1^{2})(q_2^{2}-m_2^{2})(q_3^{2}-m_3^{2})}\mathcal{F}(q_{i}^{2})\,,\\
  i \mathcal{M}_{2}(\mathcal{B}^{*},\mathcal{D}^{*},\mathcal{D}) =& \int \frac{d^{4}q_{3}}{(2\pi)^{4}} \frac{g_{B_{c2}^{*}\mathcal{B}^{*}\mathcal{D}^{*}} g_{B_{c}\mathcal{B}^{*}\mathcal{D}}g_{\mathcal{D}^{*}\mathcal{D}\mathcal{P}}B_{\mu \alpha}\qty(g^{\mu\nu}-\frac{q_{1}^{\mu}q_{1}^{\nu}}{m_1^{2}})\qty(g^{\alpha\beta}-\frac{q_{2}^{\alpha}q_{2}^{\beta}}{m_2^{2}})q_{3\nu}p_{3\beta}}{(q_1^{2}-m_1^{2})(q_2^{2}-m_2^{2})(q_3^{2}-m_3^{2})}\mathcal{F}(q_{i}^{2})  \,,\\
  i \mathcal{M}_{2}(\mathcal{B}^{*},\mathcal{D}^{*},\mathcal{D}^{*})=&\int \frac{d^{4}q_{3}}{(2\pi)^{4}} \frac{ g_{B_{c2}^{*}\mathcal{B}^{*} \mathcal{D}^{*}}g_{B_{c}\mathcal{B}^{*}\mathcal{D}^{*}}g_{\mathcal{D}^{*}\mathcal{D}^{*}\mathcal{P}}B_{\mu \alpha}\qty(g^{\mu\nu}-\frac{q_{1}^{\mu}q_{1}^{\nu}}{m_1^{2}})\qty(g^{ \alpha \beta}-\frac{q_{2}^{\alpha}q_{2}^{\beta}}{m_2^{2}})}{(q_1^{2}-m_1^{2})(q_2^{2}-m_2^{2})(q_3^{2}-m_3^{2})}\\&\times\qty(g^{\rho\sigma}-\frac{q_{3}^{\rho}q_{3}^{\sigma}}{m^{2}_{3}})  \varepsilon_{\nu \delta \rho \kappa}q_{1}^{\delta}q_{3}^{\kappa} \varepsilon_{\beta \lambda \sigma\tau}q_{3}^{\lambda}q_{2}^{\tau}\mathcal{F}(q_{i}^{2}) \,,\\
   i \mathcal{M}_{1^{(\prime)}}(\mathcal{B},\mathcal{D}^{*},\mathcal{D})=&\int \frac{d^{4}q_{3}}{(2\pi)^{4}}\frac{g_{B_{c1}^{(\prime)}\mathcal{B}\mathcal{D}^{*}}g_{B_{c}^{*}\mathcal{B}\mathcal{D}} g_{\mathcal{D}^{*}\mathcal{D}\mathcal{P}}\varepsilon^{i}_{\mu}\qty(g^{\mu \nu}-\frac{q_{2}^{\mu}q_{2}^{\nu}}{m_{2}^{2}})\varepsilon^{f}_{\rho}q_{1}^{\rho}p_{3\nu}}{(q_1^{2}-m_1^{2})(q_2^{2}-m_2^{2})(q_3^{2}-m_3^{2})}\mathcal{F}(q_{i}^{2})\,,
\\
   i \mathcal{M}_{1^{(\prime)}}(\mathcal{B}^{*},\mathcal{D}^{*},\mathcal{D})
  =&\int \frac{d^{4}q_{3}}{(2\pi)^{4}}\frac{g_{B_{c1}^{(\prime)}\mathcal{B}^{*}\mathcal{D}^{*}}g_{B_{c}^{*}\mathcal{B}^{*}\mathcal{D}}g_{\mathcal{D}^{*}\mathcal{D}\mathcal{P}}\varepsilon_{\mu\nu \rho\sigma}p_{1}^{\mu}\varepsilon_{i}^{\nu}\qty(g^{\rho \alpha}-\frac{q_{1}^{\rho}q_{1}^{\alpha}}{m_{1}^{2}})}{(q_1^{2}-m_1^{2})(q_2^{2}-m_2^{2})(q_3^{2}-m_3^{2})}\\&\times\qty(g^{\sigma\beta}-\frac{q_{2}^{\sigma}q_{2}^{\beta}}{m_{2}^{2}})\varepsilon_{\delta \lambda \kappa \alpha}p_{2}^{\delta}  \varepsilon_{f}^{\lambda}q_{1}^{\kappa}p_{3\beta}\mathcal{F}(q_{i}^{2}) \,,
\\
   i \mathcal{M}_{1^{(\prime)}}(\mathcal{B},\mathcal{D}^{*},\mathcal{D}^{*})
   =&\int \frac{d^{4}q_{3}}{(2\pi)^{4}}\frac{g_{B_{c1}^{(\prime)}\mathcal{B}\mathcal{D}^{*}}g_{B_{c}^{*} \mathcal{B} \mathcal{D}^{*}}g_{\mathcal{D}^{*}\mathcal{D}^{*}\mathcal{P}}\varepsilon^{i}_{\mu}\qty(g^{\mu \nu}-\frac{q_{2}^{\mu}q_{2}^{\nu}}{m_{2}^{2}})\varepsilon_{\rho \sigma \alpha \beta}p_{2}^{\rho}\varepsilon_{f}^{\sigma}p_{3}^{\alpha}\qty(g^{\beta\delta}-\frac{q_{3}^{\beta}q_{3}^{\delta}}{m_{3}^{2}}) \varepsilon_{\kappa \lambda \delta \nu}q_{3}^{\kappa}p_{3}^{\lambda}}{(q_1^{2}-m_1^{2})(q_2^{2}-m_2^{2})(q_3^{2}-m_3^{2})}\mathcal{F}(q_{i}^{2})\,,
\\
   i \mathcal{M}_{1^{(\prime)}}(\mathcal{B}^{*},\mathcal{D},\mathcal{D}^{*})
   =&\int \frac{d^{4}q_{3}}{(2\pi)^{4}}\frac{g_{B_{c1}^{(\prime)}\mathcal{B}^{*}\mathcal{D}}g_{B_{c}^{*}\mathcal{B}^{*}\mathcal{D}^{*}}g_{\mathcal{D}^{*}\mathcal{D}\mathcal{P}}\varepsilon^{i}_{\mu}\qty(g^{\mu\nu}-\frac{q_{1}^{\mu}q_{1}^{\nu}}{m_{1}^{2}}) (p_{2\nu}\varepsilon^{f}_{\rho}-p_{2\rho}\varepsilon^{f}_{\nu})\qty(g^{\rho \sigma}-\frac{q_{3}^{\rho}q_{3}^{\sigma}}{m_{3}^{2}})p_{3\sigma}}{(q_1^{2}-m_1^{2})(q_2^{2}-m_2^{2})(q_3^{2}-m_3^{2})} \mathcal{F}(q_{i}^{2})\,,
\\
   i \mathcal{M}_{1^{(\prime)}}(\mathcal{B}^{*},\mathcal{D}^{*},\mathcal{D}^{*})
   =&\int \frac{d^{4}q_{3}}{(2\pi)^{4}}\frac{ g_{B_{c1}^{(\prime)}\mathcal{B}^{*}\mathcal{D}^{*}}g_{B_{c}\mathcal{B}^{*}\mathcal{D}^{*}}g_{\mathcal{D}^{*}\mathcal{D}^{*}\mathcal{P}}\varepsilon_{\mu \nu \rho \sigma}p_{1}^{\mu}\varepsilon_{i}^{\nu}\qty(g^{\rho\alpha}-\frac{q_{1}^{\rho}q_{1}^{\alpha}}{m_{1}^{2}})\qty(g^{\sigma\beta}-\frac{q_{2}^{\sigma}q_{2}^{\beta}}{m_{2}^{2}})}{(q_1^{2}-m_1^{2})(q_2^{2}-m_2^{2})(q_3^{2}-m_3^{2})}\\&\times (p_{2\alpha}\varepsilon^{f}_{\delta}-p_{2\delta}\varepsilon^{f}_{\alpha})\qty(g^{\delta \lambda }-\frac{q_{3}^{\delta}q_{3}^{\lambda}}{m_{3}^{2}}) \varepsilon_{\kappa \tau \beta \lambda}p_{3}^{\kappa}q_{3}^{\tau} \mathcal{F}(q_{i}^{2}) \,.
\end{align}

In the above equations $\mathcal{F}(q_{i}^{2})$ is 
a form factor adopted for cutting off the ultraviolet divergence in the loop integrals, and has the following form
 \begin{equation}\label{formfactor}
  \mathcal{F}(q_{i}^{2})=\prod_{i}\qty(\frac{\Lambda_{i}^{2}-m_i^{2}}{\Lambda_{i}^{2}-q_{i}^{2}}),
  \end{equation} 
where $\Lambda_{i}\equiv m_{i}+\alpha \Lambda_{\text{QCD}}$ with the $m_i$ the mass of the $i-$th internal particle, and the QCD energy scale $\Lambda_{\text{QCD}}=220\, \mathrm{MeV}$ with $\alpha=1\sim 2$ as the cutoff parameter~\cite{Guo:2010ak,Cao:2023csx,Wang:2025fzj,Cao:2023gfv}. 

\begin{figure}[htbp]
\centering 
\subfigtopskip=2pt \subfigbottomskip=2pt \subfigcapskip=-10pt 
\subfigure[]{\label{fig2-a}\includegraphics[width=0.32\textwidth]{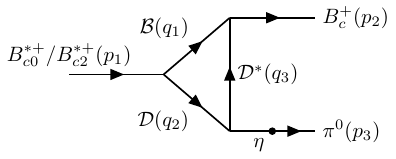}}
\subfigure[]{\label{fig2-b}\includegraphics[width=0.32\textwidth]{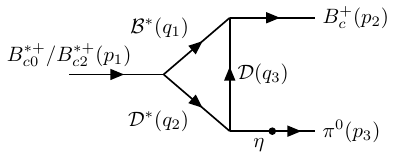}}
\subfigure[]{\label{fig2-c}\includegraphics[width=0.32\textwidth]{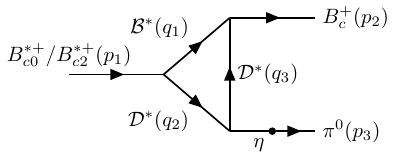}}
\subfigure[]{\label{fig2-d}\includegraphics[width=0.32\textwidth]{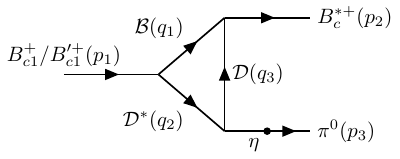}}
\subfigure[]{\label{fig2-e}\includegraphics[width=0.32\textwidth]{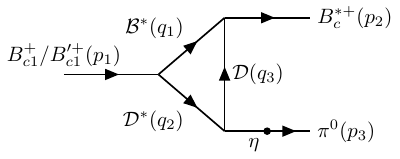}}
\subfigure[]{\label{fig2-f}\includegraphics[width=0.32\textwidth]{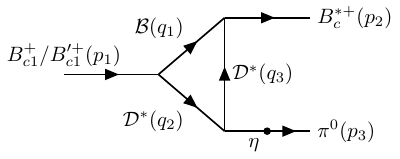}}
\subfigure[]{\label{fig2-g}\includegraphics[width=0.32\textwidth]{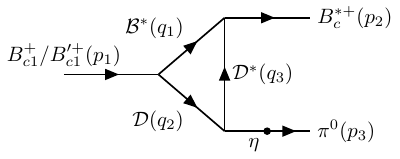}}
\subfigure[]{\label{fig2-h}\includegraphics[width=0.32\textwidth]{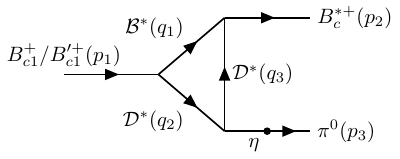}}
\caption{Schematic diagrams of the decay $B_{c}(1P)^{+}\to B_{c}^{(*)+}\pi^{0}$ via intermediate meson loops and $\eta-\pi^{0}$ mixing, where $\mathcal{B}=(B^{+},B^{0},B_{s}^{0})$ and $\mathcal{D}=(D^{0},D^{+},D_{s}^{+})$. (a)--(h) illustrate different intermediate meson loops.}\label{fig:pi_eta}
\end{figure}
For the loop diagrams involving the $\eta-\pi^0$ mixing corresponding to \cref{fig:pi_eta}, the difference from the loop diagrams of Fig.\,\ref{fig:pi} lies in the coupling constants associated with the $\eta$ and $\pi^{0}$ mesons. Specifically, for these two isospin-related channels, their amplitudes will interfere constructively in the $B_{c}(1P)^{+}\to B_{c}^{(*)+}\eta$ transition, but destructively in the $B_{c}(1P)^{+}\to B_{c}^{(*)+}\pi^{0}$ transition. In addition, for the $B_{c}(1P)^{+}\to B_{c}^{(*)+}\eta$ process, we also need to consider the contributions from the $B_{s}^{(*)}D_{s}^{(*)}D_{s}^{(*)}$ loop diagrams. We denote the amplitude as $i \mathcal{M}_{0,2,1,1^{\prime}}(P_1,P_2,P_3,\eta)$, and for each process in Fig.\,\ref{fig:pi_eta}, their explicit expressions are given by
\begin{equation}
\begin{aligned}
  i \mathcal{M}_{k}(\mathcal{B}^{(*)},\mathcal{D}^{(*)},\mathcal{D}^{(*)},\eta)=& i \mathcal{M}_{k}(\mathcal{B}^{(*)},\mathcal{D}^{(*)},\mathcal{D}^{(*)})\frac{g_{\mathcal{D}^{(*)}\mathcal{D}^{(*)}\eta}\theta_{\eta \pi^{0}}}{g_{\mathcal{D}^{(*)}\mathcal{D}^{(*)}\pi}}\,,\quad  k=0,2,1,1^{\prime}.
\end{aligned}
\end{equation}   

From the above amplitudes, we can obtain the total loop amplitude by summing up the contributions from the different loop diagrams. The decay $B_{c0}^{*+}\to B_{c}^{+}\pi^0$ is a $SPP$ type decay process, the decay $B_{c2}^{*+}\to B_{c}^{+}\pi^0$ is a $TPP$ type decay process , and the decay $B_{c1}^{(\prime)+}\to B_{c}^{*+}\pi^{0}$ is a $AVP$ type decay process so we can always parametrize the total amplitude as
 \begin{equation}
   i \mathcal{M}_{B_{c0}^{*+}}= ig_{B_{c0}^{*+}B_{c}^{+}\pi^{0}}m_{B_{c0}^{*}}\,,\quad i\mathcal{M}_{B_{c2}^{*+}}=i \frac{g_{B_{c2}^{*+}B_{c}^{+}\pi^{0}} }{m_{B_{c2}^{*}}}B_{\mu\nu}p_{2}^{\mu}p_{3}^{\nu}\,,\quad i \mathcal{M}_{B_{c1}^{(\prime)+}}=i g_{B_{c1}^{(\prime)+}B_{c}^{*+}\pi^{0}}m_{B_{c1}^{(\prime)+}}\varepsilon^{i}_{\mu}\varepsilon^{f\mu} \,.
  \end{equation} 
Then, the corresponding partial decay width is
\begin{equation}
 \begin{aligned}
  \Gamma(B_{c0}^{*+}\to B_{c}^{+}&\pi^{0})=\frac{g_{B_{c0}^{*+}B_{c}^{+}\pi^{0}}^{2}\abs{\mathbf{p}_{\pi}}}{8\pi}\,,\quad     \Gamma(B_{c2}^{*+}\to B_{c}^{+}\pi^{0})=\frac{g^{2}_{B_{c2}^{*+}B_{c}^{+}\pi^{0}}\abs{\mathbf{p}_{\pi}}^{5}}{60 \pi m_{B_{c2}^{*+}}^{4}}\,,\, \\
  &\Gamma(B_{c1}^{(\prime)+}\to B_{c}^{*+}\pi^{0})=\qty(3+\frac{\abs{\mathbf{p}_{\pi}}^{2}}{m_{B_{c}^{*+}}^{2}})\frac{g^{2}_{B_{c1}^{(\prime)+}B_{c}^{*+}\pi^{0}}\abs{\mathbf{p}_{\pi}}}{24\pi}\,, \\
 \end{aligned}
\end{equation}
where $\abs{\mathbf{p}_{\pi}}$ is the momentum of the $\pi^{0}$ in the rest frame of the initial particle.
 \subsection{Coupling constants of $B_{c(0,1,2)}^{(*)}\mathcal{B}^{(*)}\mathcal{D}^{(*)}$ vertices}
We calculate the amplitudes for the processes $B_{c(0,1,2)}^{(*)}\to \mathcal{B}^{(*)}\mathcal{D}^{(*)}$ using the $^{3}P_{0}$ model, and match them with those obtained from the effective Lagrangian approach. In this way, the coupling constants in the effective Lagrangian can be determined.

 The corresponding transition operator in the $^{3}P_{0}$ model is given by\cite{Ackleh:1996yt,Blundell:1996as,LeYaouanc:1972vsx,Micu:1968mk,LeYaouanc:1988fx} 
  \begin{equation}
    T = -3 \gamma \sum_{m} \innerproduct{1,m;1,-m}{0,0} \int d^{3}\mathbf{p}_{3} d^{3}\mathbf{p}_{4} \, \delta^{3}(\mathbf{p}_{3}+\mathbf{p}_{4}) \, \mathcal{Y}_{1}^{m}\qty(\frac{\mathbf{p}_{3}-\mathbf{p}_{4}}{2}) \chi^{34}_{1,-m} \phi^{34}_{0} \omega^{34}_{0} b^{\dagger}_{3i}(\mathbf{p}_{3}) d^{\dagger}_{4j}(\mathbf{p}_{4}),
  \end{equation}
  where $i$ and $j$ are color indices, $\chi_{1,-m}^{34}$ is the spin wave function, $\phi_{0}^{34}$ is the flavor singlet wave function, $\omega^{34}_{0} = \delta_{ij}/\sqrt{3}$ is the color singlet wave function, and $\mathcal{Y}^{m}_{1}(\mathbf{p}) = |\mathbf{p}| Y^{m}_{1}(\theta,\phi)$. 
  
  The meson wave function is given by~\cite{Hayne:1981zy}
\begin{equation}
   \begin{aligned}
   \ket{M(\mathbf{P},J,J_{z})}=&\sum_{S_{z},L_{z},c_{i}}\innerproduct{L,L_{z};S,S_{z}}{J,J_{z}}\int d^{3}\mathbf{p}_{1} d^{3}\mathbf{p}_{2}\,\delta^{3}(\mathbf{p}_{1}+\mathbf{p}_{2}-\mathbf{P})\,\psi_{N,L,L_{z}}(\mathbf{p}_{1},\mathbf{p}_{2}) \\
   &\times \frac{\delta_{c_1c_2}}{\sqrt{3}}\phi_{f_1,f_2}\chi_{s_1,s_2}^{S,S_{z}}\,b^{\dagger}_{c_1,f_1,s_1,\mathbf{p}_{1}}\,d^{\dagger}_{c_2,f_2,s_2,\mathbf{p}_{2}}\ket{0},
   \end{aligned}\label{eq:mesonwf}
 \end{equation} 
where $c_j$, $s_j$, and $f_j$ ($j=1,2$) denote the color, spin, and flavor indices of the quark, respectively. $b^{\dagger}$ and $d^{\dagger}$ are the creation operators for quarks and antiquarks, and $\psi_{N,L,L_{z}}$ is the spatial wave function in the harmonic oscillator basis.

The spatial wave functions for $S$-wave and $P$-wave mesons are given by
\begin{equation}
 \begin{aligned}
   \psi_{0,0,0}(\mathbf{p}_{1},\mathbf{p}_{2})=&\frac{1}{\pi^{3/4}R^{3/2}}\exp(-\frac{(\mathbf{p}_{1}-\mathbf{p}_{2})^{2}}{8R^{2}})\,,\\
      \psi_{0,1,m}(\mathbf{p}_{1},\mathbf{p}_{2})=&\sqrt{\frac{2}{3}}\frac{|\mathbf{p}_{1}-\mathbf{p}_{2}|}{\pi^{1/4}R^{5/2}}Y_{1}^{m}(\theta,\phi)\exp(-\frac{(\mathbf{p}_{1}-\mathbf{p}_{2})^{2}}{8R^{2}})\,, \\
 \end{aligned}
    \end{equation} 
where $Y_{1}^{m}$ is the spherical harmonic function and $R$ is the parameter of the meson wave function.

The $B_{c}(1P)^{+}$ multiplet which contains four physical states: $B_{c0}^{*+}(1^{3}P_{0})$, $B_{c2}^{*+}(1^{3}P_{2})$, $B_{c1}^{+}(1P_{1})$, and $B_{c1}^{\prime+}(1P_{1}^{\prime})$. The $B_{c1}^{+}$ and $B_{c1}^{\prime+}$ states are mixtures of the $1^{1}P_{1}$ and $1^{3}P_{1}$ configurations, with the mixing defined as
\begin{equation}
  \mqty(B_{c1}^{\prime+}\\B_{c1}^{+})=\mqty(\cos \theta_{1P}& \sin \theta_{1P}\\-\sin \theta_{1P}& \cos \theta_{1P})\mqty(1^{1}P_{1}\\ 1^{3}P_{1})\,,
 \end{equation}
where $\theta_{1P}=22.4^{\circ}$. Therefore, for $B_{c1}^{(\prime)+}$, the mixing of the wave functions must be taken into account.

The transition amplitude for $A\to BC$ in the $^{3}P_{0}$ model is given by
\begin{equation}
   \mathcal{M}_{q} = \mel{BC}{T}{A}\label{eq:ampa},
\end{equation}
and the amplitude obtained from the effective Lagrangian approach is denoted as $\mathcal{M}_{h}$. The relation between the two amplitudes is $\mathcal{M}_{h} = 8\pi^{3/2} \sqrt{m_{A}m_{B} m_{C}  }\mathcal{M}_{q}$, from which the coupling constants in the effective Lagrangian can be determined. By substituting the wave functions from Eq.\,\eqref{eq:mesonwf} into Eq.\,\eqref{eq:ampa}, and comparing the resulting amplitude with that obtained from the effective Lagrangian in Eq.\,\eqref{eq:eflbc}, the coupling constants for the $B_{c}^{(*)}$ meson vertices can be extracted. The explicit expressions for the coupling constants are given in the Appendix~\ref{app:coupling}.

\section{Numerical Results and Discussions}\label{sec:results}
Proceeding to the numerical calculations of the loop contributions for $B_{c}(1P)^{+}\to B_{c}^{(*)+}\pi^{0}$, the relevant coupling constants can be calculated using \cref{eq:gddp,eq:gbcbd,eq:gbc2bd}. The parameters used in the calculations are summarized in \cref{tab:hos} and the $^{3}P_{0}$ model coupling constant $\gamma_{c \bar{b}}=2.145$~\cite{Segovia:2012cd}~\footnote{Our value of $\gamma$ is higher than that used in~\cite{Segovia:2012cd} by a factor of $\sqrt{24\pi}$ due to the different operator convention and the different normalization of the wave function.}. The masses of the relevant particles are taken from the PDG~\cite{ParticleDataGroup:2024cfk}, the masses of $B_{c0}^{*+}$ and $B_{c2}^{*+}$ are given by~\cite{LHCb:2025uce,LHCb:2025ubr}
\begin{equation}
  m_{B_{c0}^{*+}}=6704.8\pm 5.5 \pm 2.8 \pm 0.3~\mathrm{MeV},\quad m_{B_{c2}^{*+}}=6752.4\pm 9.5 \pm 3.1 \pm 0.3~ \mathrm{MeV}\,,
 \end{equation} 
 and the masses of $B_{c1}^{+}$, $B_{c1}^{\prime+}$ and $B_{c}^{*+}$ are adopted from Ref.~\cite{Godfrey:2004ya}:
$m_{B_{c1}^{+}}=6741~\mathrm{MeV}, \ m_{B_{c1}^{\prime+}}=6750~\mathrm{MeV}$, and $m_{B_{c}^{*+}}=6338~ \mathrm{MeV}$.

The explicit values of the coupling constants are given in \cref{DDpcoupling,BcBDcoupling,BcsBDcoupling,Bc1BDcoupling,Bc11BDcoupling}.
\begin{table}[!h]
      \centering\caption{Harmonic oscillator strengths for meson wavefunctions involved in the $B_{c(0,1,2)}^{(*)}\mathcal{B}^{(*)}\mathcal{D}^{(*)}$ couplings are adopted from potential quark model studies~\cite{Godfrey:1985xj,Kokoski:1985is,Godfrey:2016nwn,Godfrey:2004ya,Godfrey:1986wj,Godfrey:2015dva}.}
      \begin{ruledtabular}
        \begin{tabular}{cccccccccccccc}
          HO Strength&$R_{D}$  &$R_{D^{*}}$     &$R_{D_{s}}$     &$R_{D^{*}_{s}}$ &$R_{B}$  &$R_{B^{*}}$&$R_{B_{s}}$&$R_{B_{s}^{*}}$&$R_{B_{c}}$&$R_{B_{c0}^{*}}$&$R_{B_{c2}^{*}}$&$R_{B_{c1}^{1}(^1P_{1})}$&$R_{B_{c1}^{3}(^{3}P_{1})}$\\\hline
          Value ($\mathrm{MeV}$)&601&516&651&562&580&542&636&595&1010&760&670&700&710
        \end{tabular}\label{tab:hos}
      \end{ruledtabular}
       \end{table}
\begin{table}[!ht]
     \centering\caption{Values of the $\mathcal{D}^{(*)}\mathcal{D}^{(*)}\mathcal{P}$ coupling constants.}
     \begin{ruledtabular}   
     \begin{tabular}{cccccc}
          Coupling constant &$g_{D^{*0}D^{*0}\pi^{0}}$&$g_{D^{*+}D^{*+}\pi^{0}}$& $g_{D^{0}D^{*0}\pi^{0}}$&$g_{D^{+}D^{*+}\pi^{0}}$&$g_{D^{*0}D^{*0}\eta}$\\\hline
     Value & $ 6.32~\mathrm{GeV}^{-1}$&$-6.32~\mathrm{GeV}^{-1}$&$12.23$&$-12.23$&$6.16~\mathrm{GeV}^{-1}$\\\hline
      Coupling constant&$g_{D^{*+}D^{*+}\eta}$& $g_{D^{0}D^{*0}\eta}$&$g_{D^{+}D^{*+}\eta}$ &$g_{D_{s}^{*+}D^{*+}_{s}\eta}$&$g_{D_{s}^{*+}D^{+}_{s}\eta}$\\\hline 
      Value&$6.16~\mathrm{GeV}^{-1}$&$11.94$&$11.94$& $-1.49~\mathrm{GeV}^{-1}$&$-3.03 $
     \end{tabular}\label{DDpcoupling}
   \end{ruledtabular}
   \end{table}

   \begin{table}[!ht]
     \centering\caption{Values of the $B_{c}^{*+}\mathcal{B}^{(*)}\mathcal{D}^{(*)}$ coupling constants.}
     \begin{ruledtabular}   
     \begin{tabular}{ccccccccc}
          Coupling constant &$g_{B_{c}^{*+}B^{+}D^{0}}$&$g_{B_{c}^{*+}B^{*+}D^{0}}$&$g_{B_{c}^{*+}B^{+}D^{*0}}$&$g_{B_{c}^{*+}B^{*+}D^{*0}}$&$g_{B_{c}^{*+}B^{0}_{s}D^{+}_{s}}$&$g_{B_{c}^{*+}B_{s}^{*0}D_{s}^{+}}$&$g_{B_{c}^{*+}B_{s}^{0}D_{s}^{*+}}$&$g_{B_{c}^{*+}B_{s}^{*0}D_{s}^{*+}}$\\\hline
     Value&$27.1$&$4.31~\mathrm{GeV}^{-1}$&$4.49~\mathrm{GeV}^{-1}$&$14.5$&$27.4$&$4.37~\mathrm{GeV}^{-1}$&$4.54~\mathrm{GeV}^{-1}$&$14.7$
     \end{tabular}
   \end{ruledtabular}\label{BcsBDcoupling}
   \end{table}
     \begin{table}[!ht]
     \centering\caption{Values of the $B_{c(0,2)}^{(*)+}\mathcal{B}^{(*)}\mathcal{D}^{(*)}$ coupling constants.}
     \begin{ruledtabular}   
     \begin{tabular}{cccccccc}
          Coupling constant &$g_{B_{c}^{+}B^{+}D^{*0}}$&$g_{B_{c}^{+}B^{*+}D^{0}}$&$g_{B_{c}^{+}B^{*+}D^{*0}}$&$g_{B_{c}^{+}B_{s}^{0}D^{*+}_{s}}$&$g_{B_{c}^{+}B_{s}^{*0}D_{s}^{+}}$&$g_{B_{c}^{+}B_{s}^{*0}D^{*+}_{s}}$&$g_{B_{c0}^{*+}B^{+}D^{0}}$\\\hline
     Value &$24.1$&$23.18$&$1.82~\mathrm{GeV}^{-1}$&$24.4$&$23.6$&$1.85~\mathrm{GeV}^{-1}$&$7.51~\mathrm{GeV}$\\\hline
     Coupling constant &$g_{B_{c0}^{*+}B^{*+}D^{*0}}$&$g_{B_{c0}^{*+}B_{s}^{0}D^{+}_{s}}$&$g_{B_{c0}^{*+}B_{s}^{*0}D_{s}^{*+}}$&$g_{B_{c2}^{*+}B^{+}D^{0}}$&$g_{B_{c2}^{*+}B^{*+}D^{*0}}$&$g_{B_{c2}^{*+}B_{s}^{0}D^{+}_{s}}$&$g_{B_{c2}^{*+}B_{s}^{*0}D_{s}^{*+}}$\\\hline
     Value & $4.50~ \mathrm{GeV}$&$8.64~ \mathrm{GeV}$&$5.23~ \mathrm{GeV}$&$3.58~\mathrm{GeV}^{-1}$&$0.365~\mathrm{GeV}^{-1}$&$3.53~ \mathrm{GeV}^{-1}$&$0.398~\mathrm{GeV}^{-1}$
     \end{tabular}\label{BcBDcoupling}
   \end{ruledtabular}
   \end{table}

   \begin{table}[!ht]
     \centering\caption{Values of the $B_{c1}^{+}\mathcal{B}^{(*)}\mathcal{D}^{(*)}$ coupling constants.}
     \begin{ruledtabular}   
     \begin{tabular}{ccccccc}
          Coupling constant &$g_{B_{c1}^{+}B^{+}D^{*0}}$&$g_{B_{c1}^{+}B^{*+}D^{*0}}$&$g_{B_{c1}^{+}B^{*+}D^{0}}$&$g_{B_{c1}^{+}B^{0}_{s}D^{*+}_{s}}$&$g_{B_{c1}^{+}B_{s}^{*0}D_{s}^{*+}}$&$g_{B_{c1}^{+}B_{s}^{*0}D_{s}^{ +}}$\\\hline
     Value&$9.13~\mathrm{GeV}$&$ 1.33$&$9.18~\mathrm{GeV}$&$10.1~\mathrm{GeV}$&$1.49 $&$10.1~\mathrm{GeV}$
     \end{tabular}
   \end{ruledtabular}\label{Bc1BDcoupling}
   \end{table}

      \begin{table}[!ht]
     \centering\caption{Values of the $B_{c1}^{\prime+}\mathcal{B}^{(*)}\mathcal{D}^{(*)}$ coupling constants.}
     \begin{ruledtabular}   
     \begin{tabular}{ccccccc}
          Coupling constant &$g_{B_{c1}^{\prime+}B^{+}D^{*0}}$&$g_{B_{c1}^{\prime+}B^{*+}D^{*0}}$&$g_{B_{c1}^{\prime+}B^{*+}D^{0}}$&$g_{B_{c1}^{\prime+}B^{0}_{s}D^{*+}_{s}}$&$g_{B_{c1}^{\prime+}B_{s}^{*0}D_{s}^{*+}}$&$g_{B_{c1}^{\prime+}B_{s}^{*0}D_{s}^{ +}}$\\\hline
     Value&$8.50~\mathrm{GeV}$&$1.24$&$8.19~\mathrm{GeV}$&$9.39~\mathrm{GeV}$&$1.38$&$9.40~\mathrm{GeV}$
     \end{tabular}
   \end{ruledtabular}\label{Bc11BDcoupling}
   \end{table}

    \begin{table}[!ht]
     \centering\caption{The partial decay widths of $B_{c}(1P)^{+}\to B_{c}^{(*)+}\pi^{0}$ with $\alpha=1.0$, $1.35$, $1.5$, $1.65$ and $2.0$ in unit of eV.}
       \begin{ruledtabular}
   \begin{tabular}{lccccc}
         $\alpha$&$1.0$ &$1.35$&$1.5$&$1.65$&$2.0$\\\hline
      $\Gamma(B_{c0}^{*+}\to B_{c}^{+}\pi^{0})$& $0.40$&$1.42$&$2.16$&$3.12$&$6.35$\\
      $\Gamma(B_{c2}^{*+}\to B_{c}^{+}\pi^{0})(\times 10^{-3})$&$0.95$&$3.79$&$6.06$&$9.17$&$20.7$\\
      $\Gamma(B_{c1}^{+} \to B_{c}^{*+}\pi^{0})$&$0.64$&$2.35$&$3.62$&$5.29$&$11.1$\\
      $\Gamma(B_{c1}^{\prime+}\to B_{c}^{*+}\pi^{0})$&$0.60$&$2.17$&$3.34$&$4.88$&$10.2$
           \end{tabular}             \end{ruledtabular}\label{tabresult1}
 \end{table}
 With the above coupling constants, we can calculate the partial decay widths of $B_{c}(1P)^{+}\to B_{c}^{(*)+}\pi^{0}$. In Table~\ref{tabresult1} we list the calculation results of the partial decay width of $B_{c}(1P)^{+}\to B_{c}^{(*)+}\pi^{0}$ with five typical cutoff parameter values $\alpha=1.0$, $1.35$, $1.5$, $1.65$ and $2.0$. For the typical value $\alpha=1.5 \pm 0.15$~\cite{Wang:2025fzj,Cao:2023gfv,Cao:2023csx}, we present the predicted partial decay widths of $B_{c}(1P)^{+}\to B_{c}^{(*)+}\pi^{0}$ in \cref{Bcresult}.
 It shows that the partial width $\Gamma(B_{c0}^{*+}\to B_{c}^{+}\pi^{0})$ is nearly three orders of magnitude larger than $\Gamma(B_{c2}^{*+}\to B_{c}^{+}\pi^{0})$. It indicates that the $B_{c}^{+}\pi^{0}$ decay channel can be used to distinguish  $B_{c0}^{*+}$ and $B_{c2}^{*+}$ from each other given sufficient data samples from experiment. Meanwhile, if $B_{c}^{+}$ and $B_{c}^{*+}$ can be distinguished by the measurement of $B_{c}^{*+}\to B_c^+\gamma$, it seems that these four states will then be separated. It is likely that in the $B_{c}^{+}\pi^{0}$ channel, one will only see one state, i.e. $B_{c0}^{*+}$, while in the $B_{c}^{*+}\pi^0$ channel one may be able to see $B_{c0}^{*+}$ and $B_{c2}^{*+}$. Notice that the isospin-violating decay process has indeed been significantly suppressed. There is no doubt that such a measurement would be extremely challenging. But for future establishment of these four states, such an effort should still be pursued. 

    \begin{table}[!ht]
    \centering\caption{Predicted partial decay widths of $B_{c}(1P)^{+}\to B_{c}^{(*)+}\pi^{0}$ for $\alpha=1.5\pm 0.15$.}
      \begin{ruledtabular}   
      \begin{tabular}{ccccc}
          Decay channel & $B_{c0}^{*+}\to B_{c}^{+}\pi^{0}$&$B_{c2}^{*+}\to B_{c}^{+}\pi^{0}$&$B_{c1}^{+}\to B_{c}^{*+}\pi^{0}$&$B_{c1}^{\prime +}\to B_{c}^{*+}\pi^{0}$ \\\hline
        Partial decay width& $2.2^{+0.7}_{-1.0}~\mathrm{eV}$&$(6.1^{+2.3}_{-3.1})\times 10^{-3}~\mathrm{eV}$&$3.6^{+1.3}_{-1.7}~\mathrm{eV}$&$3.3^{+1.2}_{-1.5}~\mathrm{eV}$
      \end{tabular}
   \end{ruledtabular}\label{Bcresult}
   \end{table}

\begin{figure}[!htbp]
    \centering
    \subfigure[]{\includegraphics[width=0.495\textwidth]{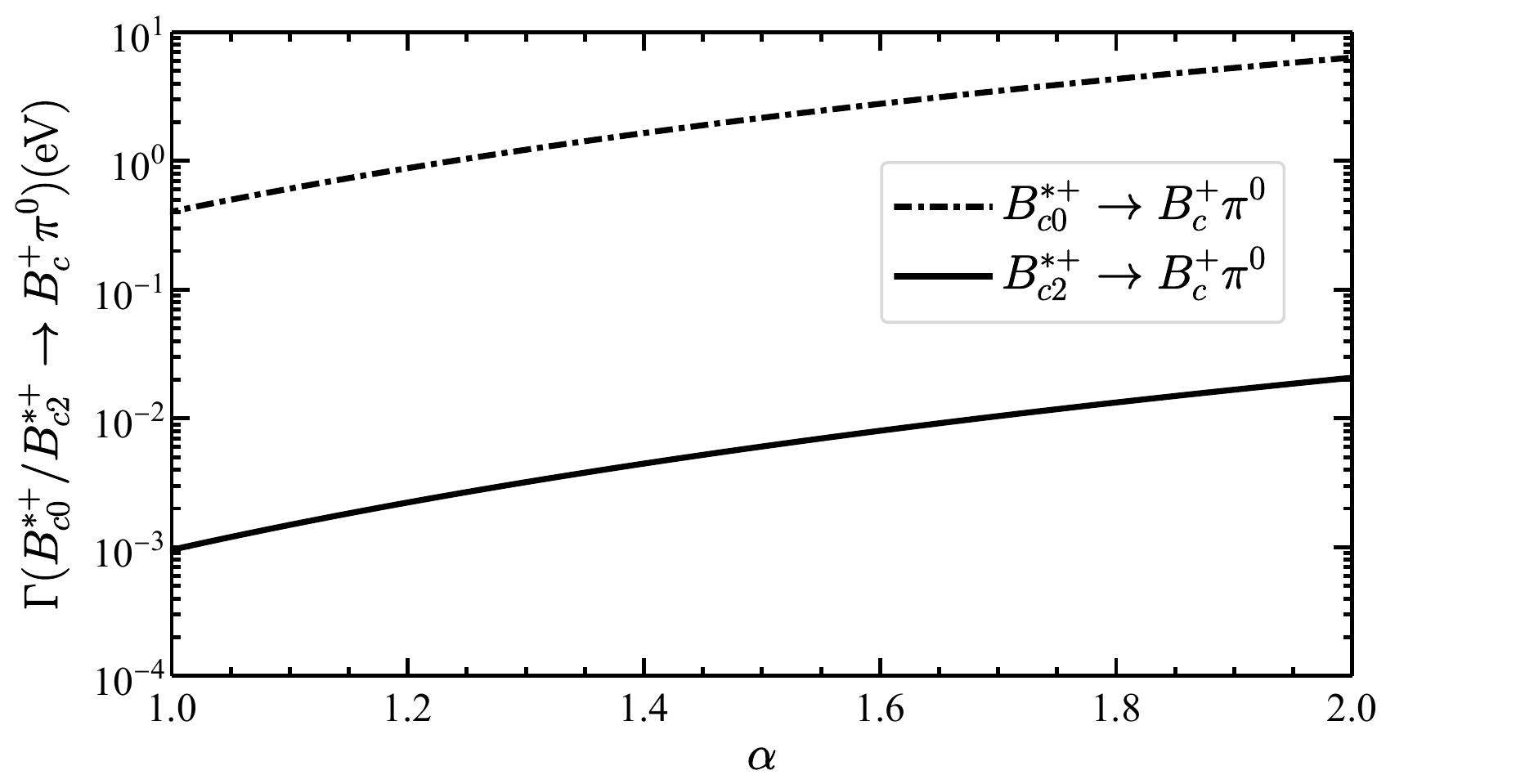}}
    \subfigure[]{\includegraphics[width=0.495\textwidth]{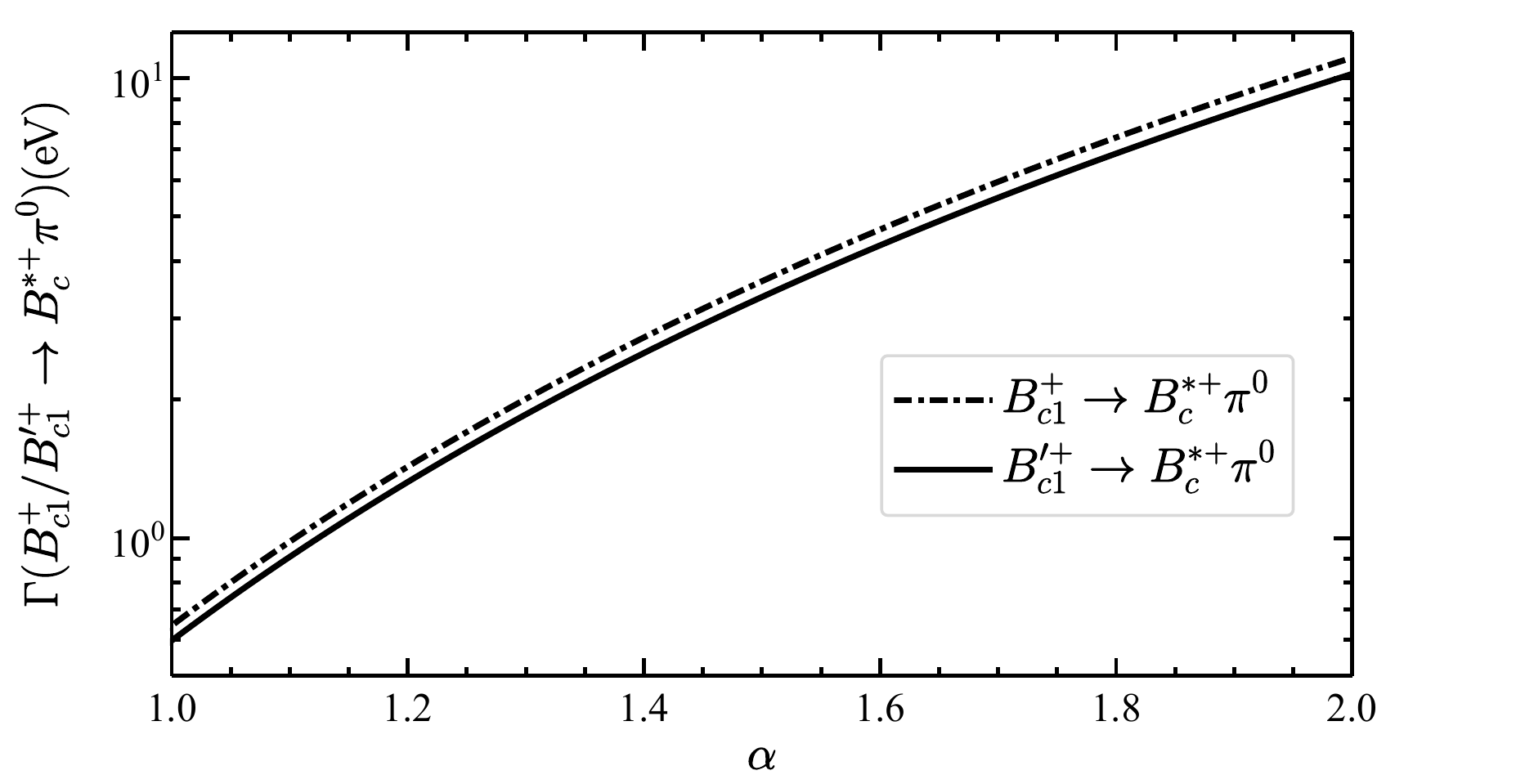}} 
    \caption{The partial decay width of $B_{c}(1P)^{+}\to B_{c}^{(*)+}\pi^{0}$ as a function of the cutoff parameter $\alpha$. }
    \label{figdecaywidth}
    \end{figure}
 \begin{figure}[!htbp]
    \centering
    \subfigure[]{\includegraphics[width=0.495\textwidth]{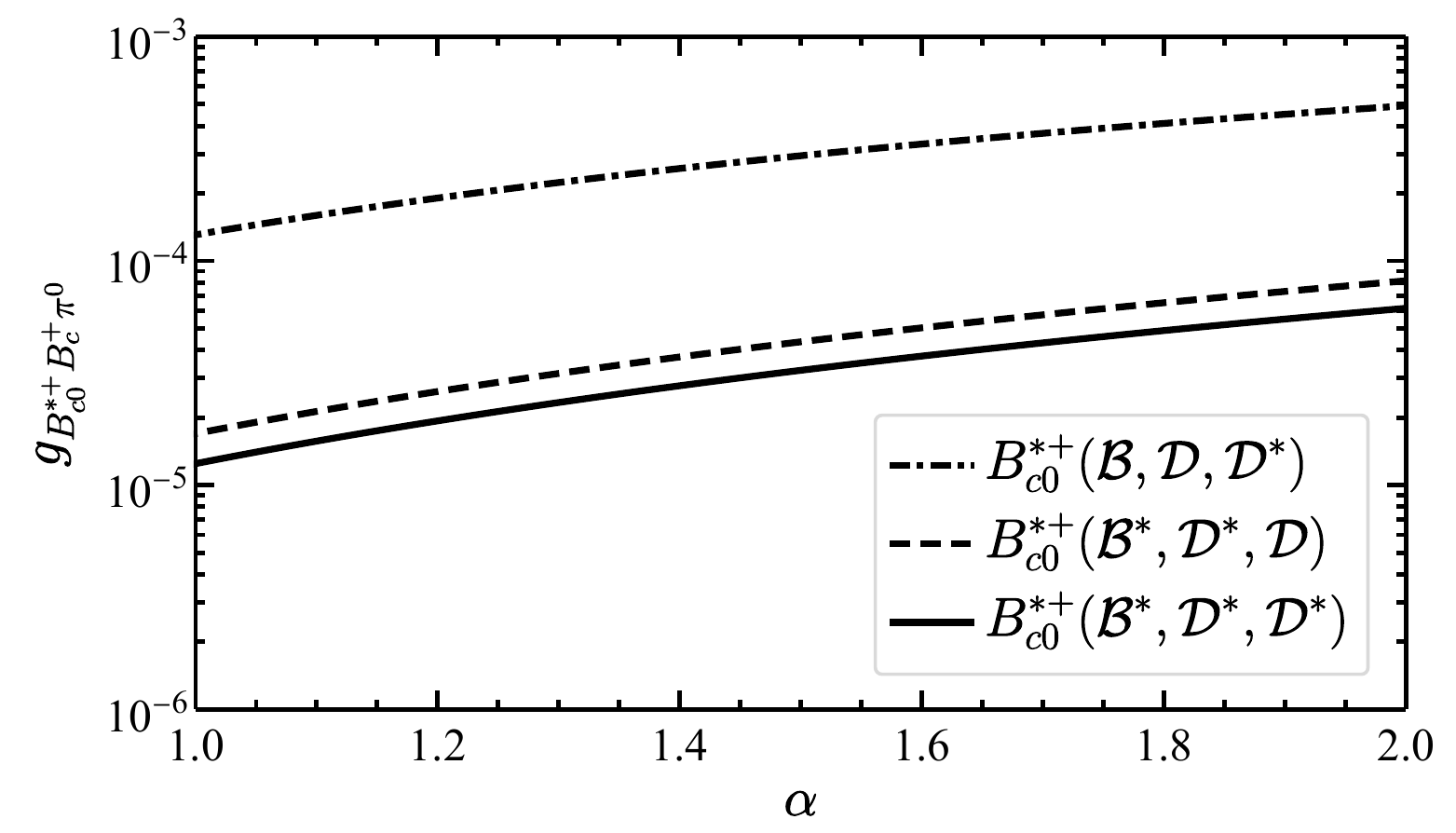}}
    \subfigure[]{\includegraphics[width=0.495\textwidth]{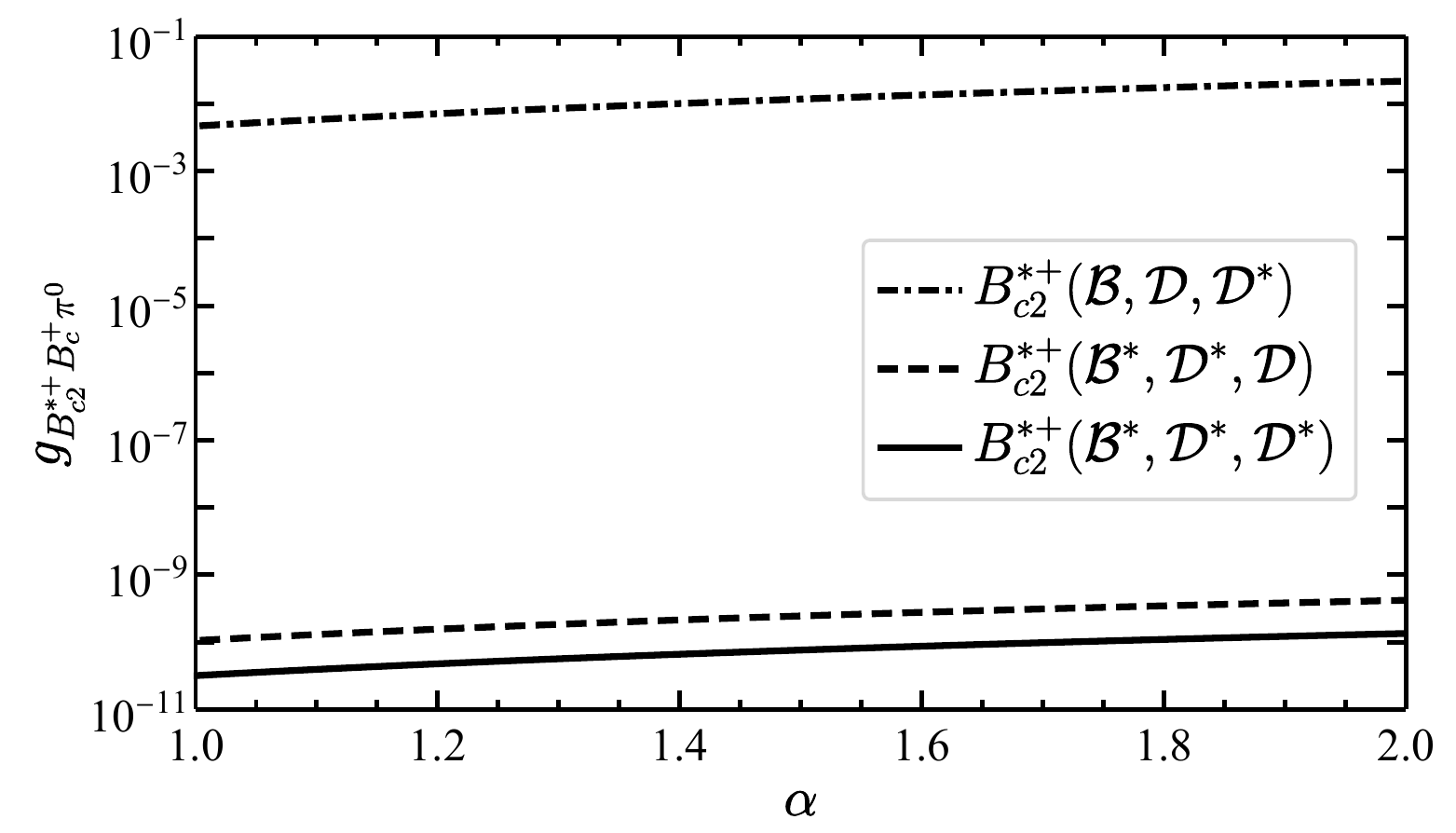}}
    \subfigure[]{\includegraphics[width=0.495\textwidth]{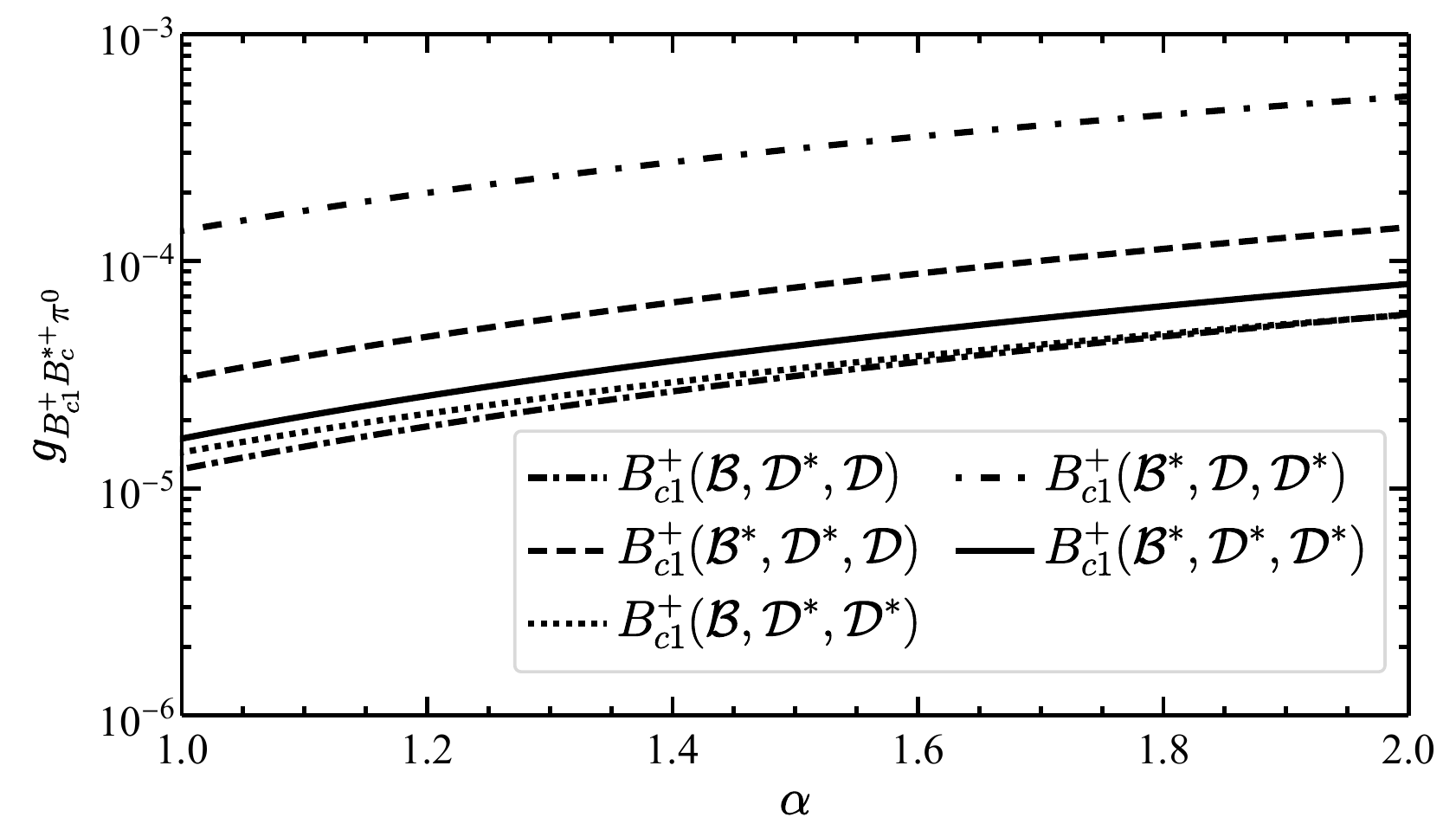}}
    \subfigure[]{\includegraphics[width=0.495\textwidth]{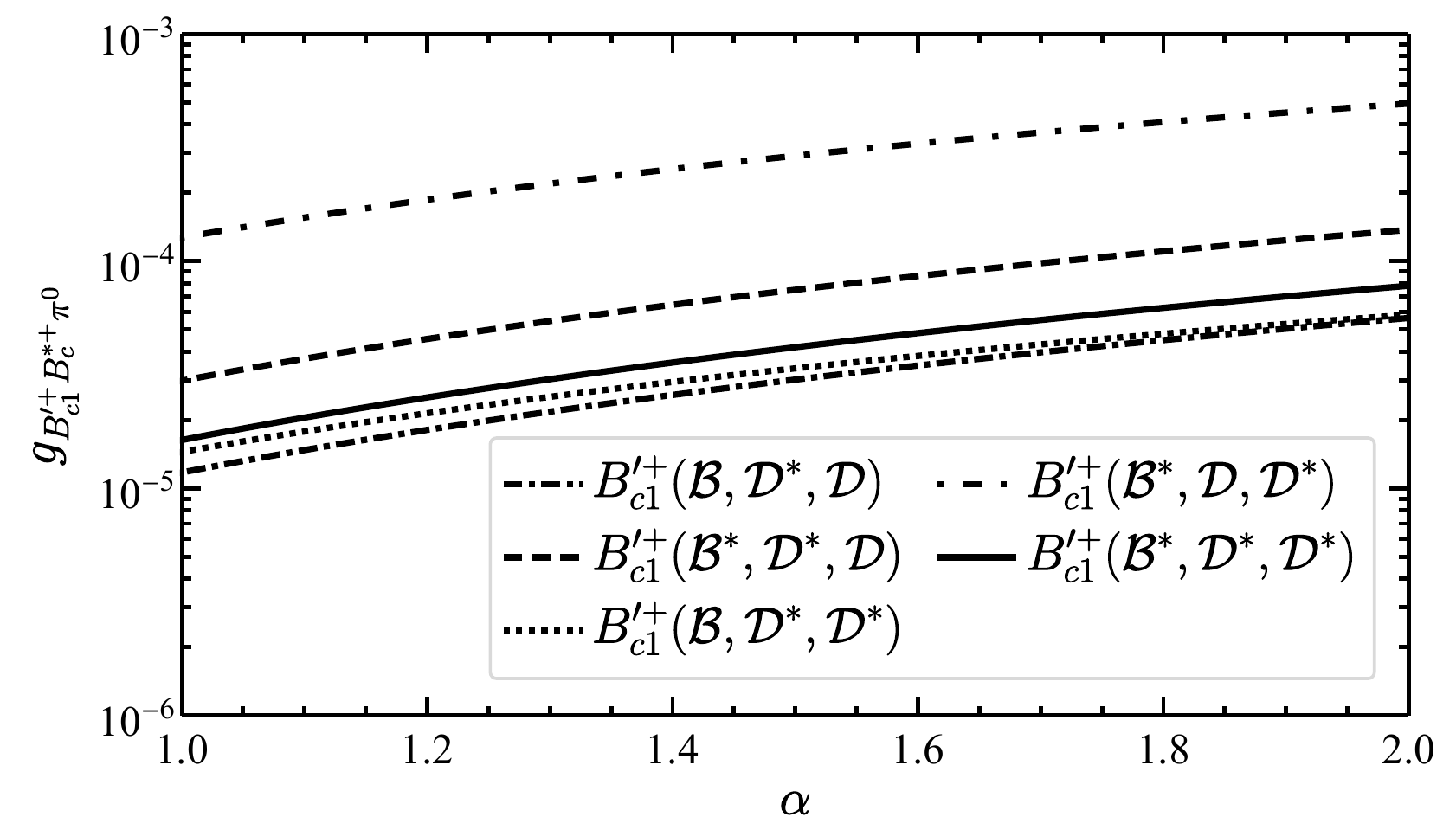}}
    \caption{Dependence of the contributions from different loop diagrams to $g_{B_{c}(1P)^{+}B_{c}^{(*)+}\pi^{0}}$ on the cutoff parameter $\alpha$. (a)$-$(d) correspond to different $B_{c}(1P)^{+}$ states.}
    \label{gcouple}
    \end{figure} 
    To see more clearly the role played by the intermediate loop transitions, we plot the dependence of the decay width of $B_{c}(1P)^{+}\to B_{c}^{(*)+}\pi^{0}$ on the cut-off parameter $\alpha$ in \cref{figdecaywidth}. The behavior of the decay widths as a function of the cutoff parameter $\alpha$ is found to be consistent with $D_{s}^{*+}\to D_{s}^{+}\pi^{0}$~\cite{Wang:2025fzj}. To further clarify the contributions from different loop diagrams in the processes $B_{c}(1P)^{+}\to B_{c}^{(*)+}\pi^{0}$, we present in \cref{gcouple} the dependence of the different loop contributions to $g_{B_{c}(1P)^{+}B_{c}^{(*)+}\pi^{0}}$ on the cutoff parameter $\alpha$. It is worth noting that for all four processes of $B_{c}(1P)^{+}\to B_{c}^{(*)+}\pi^{0}$, the loop diagrams involving the exchange of a $\mathcal{D}^{*}$ meson and containing a $\mathcal{D}$ meson as an internal line provide the dominant contribution, which is more than one order of magnitude larger than the other loop contributions. This is especially pronounced in the case of $B_{c2}^{*+}\to B_{c}^{+}\pi^{0}$.
    
\section{Summary} \label{sec:summary}
In this work we calculate the partial decay widths for the four isospin-violating decay channels $B_{c}(1P)^{+}\to B_{c}^{(*)+}\pi^{0}$ using the effective Lagrangian approach. We assume that the isospin-violating $\pi^0$ production via the U(1) anomaly term is dual to the intermediate meson loops involving $\mathcal{B}^{(*)}$ and $\mathcal{D}^{(*)}$ rescatterings. The isospin-violating transition is a consequence of the cancellations between the two kinds of loops involving the intermediate heavy mesons containing $u\bar{u}$ or $d\bar{d}$. In particular, we find that the loop diagrams with $\mathcal{B}^{(*)}\mathcal{D}$ rescatterings by exchanging a $\mathcal{D}^{*}$ provide the dominant contributions. Our results show that the partial decay width of $B_{c0}^{*+}\to B_{c}^{+}\pi^{0}$ is about three orders of magnitude larger than that for $B_{c2}^{*+}\to B_{c}^{+}\pi^{0}$. This significant difference can be exploited experimentally to distinguish between the $B_{c0}^{*+}$ and $B_{c2}^{*+}$ states in the final state of $B_{c}^{+}\pi^{0}$, while the two axial-vector states $B_{c1}^{+}/B_{c1}'^{+}$ can be possibly identified in $B_{c1}^{+}/B_{c1}'^{+}\to B_{c}^{*+}\pi^{0}$. Although these measurements would be challenging for the present and even future experiment, we emphasize that they are crucial for finally establishing these states in experiment. 

\begin{acknowledgements}
Useful discussions with Dr. Liu-Pan An are acknowledged. This work is supported, in part, by the National Natural Science Foundation of China (Grant No. 12235018), DFG and NSFC funds to the Sino-German CRC 110 ``Symmetries and the Emergence of Structure in QCD" (NSFC Grant No. 12070131001, DFG Project-ID 196253076), National Key Basic Research Program of China under Contract No. 2020YFA0406300, and Strategic Priority Research Program of Chinese Academy of Sciences (Grant No. XDB34030302).
\end{acknowledgements}

\appendix
\section{Coupling constants of bottom-charm mesons}\label{app:coupling}
Explicit expressions for the $B_{c(0,1,2)}^{(*)}\mathcal{B}^{(*)}\mathcal{D}^{(*)}$ couplings, which are calculated in the $^3P_0$ model, are listed below. Namely, for the $B_c^{+}$ couplings to $\mathcal{B}^{(*)}\mathcal{D}^{(*)}$, we have
\begin{equation}
  \begin{aligned}
  g_{B_{c}\mathcal{B}^{*}\mathcal{D}}=&\frac{4\sqrt{2}\pi^{\frac{1}{4}}\sqrt{m_{B_{c}}m_{\mathcal{B}^{*}}m_{\mathcal{D}}}\gamma_{c\bar{b}}(R_{B_c} R_{\mathcal{B}^*} R_{\mathcal{D}})^{3/2} (R_{B_c}^2 (R_{\mathcal{B}^*}^2+R_{\mathcal{D}}^2)+2 R_{\mathcal{B}^*}^2 R_{\mathcal{D}}^2)}{\sqrt{3}  (R_{B_c}^2 (R_{\mathcal{B}^*}^2+R_{\mathcal{D}}^2)+R_{\mathcal{B}^*}^2 R_{\mathcal{D}}^2)^{5/2}}\,,\\
  g_{B_{c}\mathcal{B} \mathcal{D}^{*}}=&\frac{4\sqrt{2}\pi^{\frac{1}{4}}\sqrt{m_{B_{c}}m_{\mathcal{B}^{*}}m_{\mathcal{D}^*}}\gamma_{c\bar{b}}(R_{B_c} R_{\mathcal{B}} R_{\mathcal{D}^*})^{3/2} (R_{B_c}^2 (R_{\mathcal{B}}^2+R_{\mathcal{D}^*}^2)+2 R_{\mathcal{B}}^2 R_{\mathcal{D}^*}^2)}{\sqrt{3}  (R_{B_c}^2 (R_{\mathcal{B}}^2+R_{\mathcal{D}^*}^2)+R_{\mathcal{B}}^2 R_{\mathcal{D}^*}^2)^{5/2}}\,,\\
  g_{B_{c}\mathcal{B}^{*} \mathcal{D}^{*}}=&\frac{4\sqrt{2}\pi^{\frac{1}{4}}\sqrt{m_{\mathcal{B}^{*}}m_{\mathcal{D}^{*}}}\gamma_{c\bar{b}}(R_{B_c} R_{\mathcal{B}^*} R_{\mathcal{D}^*})^{3/2} (R_{B_c}^2 (R_{\mathcal{B}^*}^2+R_{\mathcal{D}^*}^2)+2 R_{\mathcal{B}^*}^2 R_{\mathcal{D}^*}^2)}{\sqrt{3m_{B_{c}}}(R_{B_c}^2 (R_{\mathcal{B}^*}^2+R_{\mathcal{D}^*}^2)+R_{\mathcal{B}^*}^2 R_{\mathcal{D}^*}^2)^{5/2}} \ .
  \end{aligned}\label{eq:gbcbd}
 \end{equation} 
For the $B_c^{*+}$ couplings to $\mathcal{B}^{(*)}\mathcal{D}^{(*)}$, we have
 \begin{equation}
   \begin{aligned}
    g_{B_{c}^{*}\mathcal{B}\mathcal{D}}=&\frac{4\sqrt{2}\pi^{\frac{1}{4}}\sqrt{m_{B_{c}^{*}}m_{\mathcal{B}}m_{\mathcal{D}}}\gamma_{c\bar{b}}(R_{B_c^{*}} R_{\mathcal{B}} R_{\mathcal{D}})^{3/2} (R_{B_c^{*}}^2 (R_{\mathcal{B}}^2+R_{\mathcal{D}}^2)+2 R_{\mathcal{B}}^2 R_{\mathcal{D}}^2)}{\sqrt{3} (R_{B_c^{*}}^2 (R_{\mathcal{B}}^2+R_{\mathcal{D}}^2)+R_{\mathcal{B}}^2 R_{\mathcal{D}}^2)^{5/2}} \ , \\
     g_{B_{c}^{*}\mathcal{B}^{*}\mathcal{D}}=&\frac{4\sqrt{2}\pi^{\frac{1}{4}}\sqrt{m_{\mathcal{B}}m_{\mathcal{D}}}\gamma_{c\bar{b}}(R_{B_c^{*}} R_{\mathcal{B}^{*}} R_{\mathcal{D}})^{3/2} (R_{B_c^{*}}^2 (R_{\mathcal{B}^{*}}^2+R_{\mathcal{D}}^2)+2 R_{\mathcal{B}^{*}}^2 R_{\mathcal{D}}^2)}{\sqrt{3m_{B_{c}^{*}}}  (R_{B_c^{*}}^2 (R_{\mathcal{B}^{*}}^2+R_{\mathcal{D}}^2)+R_{\mathcal{B}^{*}}^2 R_{\mathcal{D}}^2)^{5/2}} \ ,\\
       g_{B_{c}^{*}\mathcal{B}\mathcal{D}^{*}}=&\frac{4\sqrt{2}\pi^{\frac{1}{4}}\sqrt{m_{\mathcal{B}}m_{\mathcal{D}^{*}}}\gamma_{c\bar{b}}(R_{B_c^{*}} R_{\mathcal{B}} R_{\mathcal{D}^{*}})^{3/2} (R_{B_c^{*}}^2 (R_{\mathcal{B}}^2+R_{\mathcal{D}^{*}}^2)+2 R_{\mathcal{B}}^2 R_{\mathcal{D}^{*}}^2)}{\sqrt{3m_{B_{c}^{*}}} (R_{B_c^{*}}^2 (R_{\mathcal{B}}^2+R_{\mathcal{D}^{*}}^2)+R_{\mathcal{B}}^2 R_{\mathcal{D}^{*}}^2)^{5/2}} \ ,\\
        g_{B_{c}^{*}\mathcal{B}^{*}\mathcal{D}^{*}}=&\frac{2\sqrt{2}\pi^{\frac{1}{4}}\sqrt{m_{B_{c}^{*}}m_{\mathcal{B}^{*}}m_{\mathcal{D}^{*}}}\gamma_{c\bar{b}}(R_{B_c^{*}} R_{\mathcal{B}^{*}} R_{\mathcal{D}^{*}})^{3/2} (R_{B_c^{*}}^2 (R_{\mathcal{B}^{*}}^2+R_{\mathcal{D}^{*}}^2)+2 R_{\mathcal{B}^{*}}^2 R_{\mathcal{D}^{*}}^2)}{\sqrt{3} (R_{B_c^{*}}^2 (R_{\mathcal{B}^{*}}^2+R_{\mathcal{D}^{*}}^2)+R_{\mathcal{B}^{*}}^2 R_{\mathcal{D}^{*}}^2)^{5/2}} \ .
  \end{aligned}
  \end{equation} 
For the $B_{c0}^{*+}$ couplings to $\mathcal{B}^{(*)}\mathcal{D}^{(*)}$, we have
 \begin{equation}
  \begin{aligned}
  g_{B_{c0}^{*}\mathcal{B}\mathcal{D}}=&\frac{8\pi^{\frac{1}{4}}\sqrt{m_{B_{c0}^{*}}m_{\mathcal{B}}m_{\mathcal{D}}}\gamma_{c\bar{b}}R_{\mathcal{B}} R_{\mathcal{D}} (R_{B_{c0}^{*}}R_{\mathcal{B}} R_{\mathcal{D}})^{5/2}}{(R_{B_{c0}^{*}}^2(R_{\mathcal{B}}^2+R_{\mathcal{D}}^2)+R_{\mathcal{B}}^2 R_{\mathcal{D}}^2)^{\frac{5}{2}}}\,,\\
  g_{B_{c0}^{*}\mathcal{B}^{*} \mathcal{D}^{*}}=&\frac{16\pi^{\frac{1}{4}}\sqrt{m_{B_{c0}^{*}}m_{\mathcal{B}^{*}}m_{\mathcal{D}^{*}}}\gamma_{c\bar{b}}R_{\mathcal{B}^{*}} R_{\mathcal{D}^{*}} (R_{B_{c0}^{*}}R_{\mathcal{B}^{*}} R_{\mathcal{D}^{*}})^{5/2}}{3(R_{B_{c0}^{*}}^2(R_{\mathcal{B}^{*}}^2+R_{\mathcal{D}^{*}}^2)+R_{\mathcal{B}^{*}}^2 R_{\mathcal{D}^{*}}^2)^{\frac{5}{2}}}\, .
  \end{aligned}\label{eq:gbc0bd}
 \end{equation} 
For the $B_{c2}^{*+}$ couplings to $\mathcal{B}^{(*)}\mathcal{D}^{(*)}$, we have
 \begin{equation}
  \begin{aligned}
  g_{B_{c2}^{*}\mathcal{B}\mathcal{D}}=&\frac{\pi^{\frac{1}{4}}\sqrt{m_{B_{c2}^{*}} m_{\mathcal{B}}m_{\mathcal{D}}}\gamma_{c\bar{b}}R_{B_{c2}^*}^{5/2} (R_{\mathcal{B}} R_{\mathcal{D}})^{\frac{3}{2}} (R_{\mathcal{B}}^2+R_{\mathcal{D}}^2) (R_{B_{c2}^*}^2 (R_{\mathcal{B}}^2+R_{\mathcal{D}}^2)+2 R_{\mathcal{B}}^2 R_{\mathcal{D}}^2) }{\sqrt{3} (R_{B_{c2}^*}^2 (R_{\mathcal{B}}^2+R_{\mathcal{D}}^2)+R_{\mathcal{B}}^2 R_{\mathcal{D}}^2)^{7/2}}\,,\\
  g_{B_{c2}^{*}\mathcal{B}^{*} \mathcal{D}^{*}}=&\frac{16\pi^{\frac{1}{4}}\sqrt{m_{B_{c2}^{*}}m_{\mathcal{B}^{*}}m_{\mathcal{D}^{*}}}\gamma_{c\bar{b}}R_{\mathcal{B}^{*}} R_{\mathcal{D}^{*}} (R_{B_{c2}}R_{\mathcal{B}^{*}} R_{\mathcal{D}^{*}})^{5/2}}{\sqrt{3}(R_{B_{c2}^{*}}^2(R_{\mathcal{B}^{*}}^2+R_{\mathcal{D}^{*}}^2)+R_{\mathcal{B}^{*}}^2 R_{\mathcal{D}^{*}}^2)^{\frac{5}{2}}}\,.
  \end{aligned}\label{eq:gbc2bd}
 \end{equation} 
For the $B_{c1}^{(\prime)+}$ couplings to $\mathcal{B}^{(*)}\mathcal{D}^{(*)}$, taking into account the mixing between the $1^1P_1$ and $1^3P_1$ we have
\begin{equation}
  \begin{aligned}
   g_{B_{c1}^{\prime}\mathcal{B}\mathcal{D}^{*}}=&\cos \theta_{1P}\frac{8\pi^{\frac{1}{4}}\sqrt{m_{B_{c1}^{\prime}}m_{\mathcal{B}}m_{\mathcal{D}^{*}}}\gamma_{c\bar{b}}R_{\mathcal{B}}R_{\mathcal{D}^{*}}(R_{B_{c1}^{1}}R_{\mathcal{B}}R_{\mathcal{D}^{*}})^{\frac{5}{2}}}{\sqrt{3}(R_{B_{c1}^{1}}^2 (R_{\mathcal{B}}^2+R_{\mathcal{D}^{*}}^2)+R_{\mathcal{B}}^2 R_{\mathcal{D}^{*}}^2)^{5/2}}\\
           &+\sqrt{2}\sin \theta_{1P}\frac{8\pi^{\frac{1}{4}}\sqrt{m_{B_{c1}^{\prime}}m_{\mathcal{B}}m_{\mathcal{D}^{*}}}\gamma_{c\bar{b}}R_{\mathcal{B}}R_{\mathcal{D}^{*}}(R_{B_{c1}^{3}}R_{\mathcal{B}}R_{\mathcal{D}^{*}})^{\frac{5}{2}}}{\sqrt{3}(R_{B_{c1}^{3}}^2 (R_{\mathcal{B}}^2+R_{\mathcal{D}^{*}}^2)+R_{\mathcal{B}}^2 R_{\mathcal{D}^{*}}^2)^{5/2}} \ ,\\
           g_{B_{c1}\mathcal{B}\mathcal{D}^{*}}=&-\sin  \theta_{1P}\frac{8\pi^{\frac{1}{4}}\sqrt{m_{B_{c1}}m_{\mathcal{B}}m_{\mathcal{D}^{*}}}\gamma_{c\bar{b}}R_{\mathcal{B}}R_{\mathcal{D}^{*}}(R_{B_{c1}^{1}}R_{\mathcal{B}}R_{\mathcal{D}^{*}})^{\frac{5}{2}}}{\sqrt{3}(R_{B_{c1}^{1}}^2 (R_{\mathcal{B}}^2+R_{\mathcal{D}^{*}}^2)+R_{\mathcal{B}}^2 R_{\mathcal{D}^{*}}^2)^{5/2}}\\
           &+\sqrt{2}\cos  \theta_{1P}\frac{8\pi^{\frac{1}{4}}\sqrt{m_{B_{c1}}m_{\mathcal{B}}m_{\mathcal{D}^{*}}}\gamma_{c\bar{b}}R_{\mathcal{B}}R_{\mathcal{D}^{*}}(R_{B_{c1}^{3}}R_{\mathcal{B}}R_{\mathcal{D}^{*}})^{\frac{5}{2}}}{\sqrt{3}(R_{B_{c1}^{3}}^2 (R_{\mathcal{B}}^2+R_{\mathcal{D}^{*}}^2)+R_{\mathcal{B}}^2 R_{\mathcal{D}^{*}}^2)^{5/2}} \ ,\\
  \end{aligned}
 \end{equation} 
  \begin{equation}
           \begin{aligned}
           g_{B_{c1}^{\prime}\mathcal{B}^{*}\mathcal{D}}=&\cos \theta_{1P}\frac{8\pi^{\frac{1}{4}}\sqrt{m_{B_{c1}^{\prime}}m_{\mathcal{B}^{*}}m_{\mathcal{D}}}\gamma_{c\bar{b}}R_{\mathcal{B}^{*}}R_{\mathcal{D}}(R_{B_{c1}^{1}}R_{\mathcal{B}^{*}}R_{\mathcal{D}})^{\frac{5}{2}}}{\sqrt{3}(R_{B_{c1}^{1}}^2 (R_{\mathcal{B}^{*}}^2+R_{\mathcal{D}}^2)+R_{\mathcal{B}^{*}}^2 R_{\mathcal{D}}^2)^{5/2}}\\
           &+\sqrt{2}\sin \theta_{1P}\frac{8\pi^{\frac{1}{4}}\sqrt{m_{B_{c1}^{\prime}}m_{\mathcal{B}^{*}}m_{\mathcal{D}}}\gamma_{c\bar{b}}R_{\mathcal{B}^{*}}R_{\mathcal{D}}(R_{B_{c1}^{3}}R_{\mathcal{B}^{*}}R_{\mathcal{D}})^{\frac{5}{2}}}{\sqrt{3}(R_{B_{c1}^{3}}^2 (R_{\mathcal{B}^{*}}^2+R_{\mathcal{D}}^2)+R_{\mathcal{B}^{*}}^2 R_{\mathcal{D}}^2)^{5/2}} \ ,\\
           g_{B_{c1}\mathcal{B}^{*}\mathcal{D}}=&-\sin  \theta_{1P}\frac{8\pi^{\frac{1}{4}}\sqrt{m_{B_{c1}}m_{\mathcal{B}^{*}}m_{\mathcal{D}}}\gamma_{c\bar{b}}R_{\mathcal{B}^{*}}R_{\mathcal{D}}(R_{B_{c1}^{1}}R_{\mathcal{B}^{*}}R_{\mathcal{D}})^{\frac{5}{2}}}{\sqrt{3}(R_{B_{c1}^{1}}^2 (R_{\mathcal{B}^{*}}^2+R_{\mathcal{D}}^2)+R_{\mathcal{B}^{*}}^2 R_{\mathcal{D}}^2)^{5/2}}\\
           &+\sqrt{2}\cos  \theta_{1P}\frac{8\pi^{\frac{1}{4}}\sqrt{m_{B_{c1}}m_{\mathcal{B}^{*}}m_{\mathcal{D}}}\gamma_{c\bar{b}}R_{\mathcal{B}^{*}}R_{\mathcal{D}}(R_{B_{c1}^{3}}R_{\mathcal{B}^{*}}R_{\mathcal{D}})^{\frac{5}{2}}}{\sqrt{3}(R_{B_{c1}^{3}}^2 (R_{\mathcal{B}^{*}}^2+R_{\mathcal{D}}^2)+R_{\mathcal{B}^{*}}^2 R_{\mathcal{D}}^2)^{5/2}} \ ,
           \end{aligned}
          \end{equation} 
           \begin{equation}
           \begin{aligned}
           g_{B_{c1}^{\prime}\mathcal{B}^{*}\mathcal{D}^{*}}=&\cos \theta_{1P}\frac{8\pi^{\frac{1}{4}}\sqrt{m_{\mathcal{B}^{*}}m_{\mathcal{D}^{*}}}\gamma_{c\bar{b}}R_{\mathcal{B}^{*}}R_{\mathcal{D}^{*}}(R_{B_{c1}^{1}}R_{\mathcal{B}^{*}}R_{\mathcal{D}^{*}})^{\frac{5}{2}}}{\sqrt{3m_{B_{c1}^{\prime}}}(R_{B_{c1}^{1}}^2 (R_{\mathcal{B}^{*}}^2+R_{\mathcal{D}^{*}}^2)+R_{\mathcal{B}^{*}}^2 R_{\mathcal{D}^{*}}^2)^{5/2}}\\
           &+\sqrt{2}\sin \theta_{1P}\frac{8\pi^{\frac{1}{4}}\sqrt{m_{\mathcal{B}^{*}}m_{\mathcal{D}^{*}}}\gamma_{c\bar{b}}R_{\mathcal{B}^{*}}R_{\mathcal{D}^{*}}(R_{B_{c1}^{3}}R_{\mathcal{B}^{*}}R_{\mathcal{D}^{*}})^{\frac{5}{2}}}{\sqrt{3m_{B_{c1}^{\prime}}}(R_{B_{c1}^{3}}^2 (R_{\mathcal{B}^{*}}^2+R_{\mathcal{D}^{*}}^2)+R_{\mathcal{B}^{*}}^2 R_{\mathcal{D}^{*}}^2)^{5/2}} \ ,\\
           g_{B_{c1}\mathcal{B}^{*}\mathcal{D}^{*}}=&-\sin  \theta_{1P}\frac{8\pi^{\frac{1}{4}}\sqrt{m_{\mathcal{B}^{*}}m_{\mathcal{D}^{*}}}\gamma_{c\bar{b}}R_{\mathcal{B}^{*}}R_{\mathcal{D}^{*}}(R_{B_{c1}^{1}}R_{\mathcal{B}^{*}}R_{\mathcal{D}^{*}})^{\frac{5}{2}}}{\sqrt{3m_{B_{c1}}}(R_{B_{c1}^{1}}^2 (R_{\mathcal{B}^{*}}^2+R_{\mathcal{D}^{*}}^2)+R_{\mathcal{B}^{*}}^2 R_{\mathcal{D}^{*}}^2)^{5/2}}\\
           &+\sqrt{2}\cos  \theta_{1P}\frac{8\pi^{\frac{1}{4}}\sqrt{m_{\mathcal{B}^{*}}m_{\mathcal{D}^{*}}}\gamma_{c\bar{b}}R_{\mathcal{B}^{*}}R_{\mathcal{D}^{*}}(R_{B_{c1}^{3}}R_{\mathcal{B}^{*}}R_{\mathcal{D}^{*}})^{\frac{5}{2}}}{\sqrt{3m_{B_{c1}}}(R_{B_{c1}^{3}}^2 (R_{\mathcal{B}^{*}}^2+R_{\mathcal{D}^{*}}^2)+R_{\mathcal{B}^{*}}^2 R_{\mathcal{D}^{*}}^2)^{5/2}} \ .
           \end{aligned}
          \end{equation} 

\bibliographystyle{unsrt}
\bibliography{Bc1P_decay.bib}
\end{document}